\documentclass[12pt,a4paper]{article}

\usepackage{amsmath}
\usepackage{geometry}
\usepackage{amssymb}
\usepackage{bm}
\usepackage{amsthm}
\usepackage{natbib}
\usepackage{graphicx}
\usepackage{caption,subcaption}
\usepackage{color}
\usepackage{float}
\usepackage{authblk}
\usepackage[titletoc,title]{appendix}
\usepackage{setspace}
\usepackage{diagbox}   % \slash line in Table
\usepackage{multirow}
\usepackage{array}
\usepackage{booktabs}
\usepackage{adjustbox}
\usepackage{rotating}
\usepackage{pdflscape}
\usepackage{algorithm}
\usepackage{algpseudocode}

\usepackage{longtable}
\usepackage{slashbox}   % \slash line in Table
\usepackage{tabularx}

\usepackage[markup=underlined, defaultcolor=orange]{changes}
\definechangesauthor[name={F.P.ZHANG}, color=orange]{ZFP}
\newtheorem{theorem}{Theorem}

\newcommand{\bbeta}{\bm{\beta}}

\newcommand{\bx}{\bm{x}}
\newcommand{\bX}{\bm{X}}

\newcommand{\bZ}{\bm{Z}}

\graphicspath{{figures/}}

\newcommand*{\thisdraft}{This version of draft: October 12, 2026}

\begin{document}

\def\spacingset#1{\renewcommand{\baselinestretch}%
{#1}\small\normalsize} \spacingset{1}
\spacingset{1.8}

\date{}
\date{\thisdraft}
\title{
	Conditional expectile-based value-at-risk: Semiparametric estimation and inference
}

%%%%%%%%%%%%%%%%%%%%
%%##### anonymous authors or not, 
    %% take  non-anonymous=1, anonymous=0
   \def\mycmd{1}
	
   \if\mycmd0
       \author{}
   \else
     \author[1]{Yuhan Ma} 
    \author[2]{Heng Peng
    %\thanks{Correspondence author. Email:  hpeng@hkbu.edu.hk}
    }
 	\author[1]{Feipeng Zhang
 		%\thanks{Correspondence author. Email:  zhangfp108@163.com}
 	} 
     
     \affil[1]{\small School of Economics and Finance, Xi'an Jiaotong University, Xi'an, 710049, China}

     \affil[2]{\small Department of Mathematics, Hong Kong Baptist University, Hong Kong, China}

   \fi
%%%%%%%%%%%%%%%%%%%%%%%%%%%%%%%%%%%%%%%%%%%%%%%%%%
\maketitle

%\date{%\thisdraft \\ \firstdraft}

\vspace*{-9mm}
\begin{spacing}{1.2}
\noindent 	
\textbf{Abstract:} 	
%Conditional expectile-based value-at-risk (EVaR) has been recommended in financial risk management due to its coherency and elicitability. 
Conditional expectile-based value-at-risk (EVaR) has attracted increasing attention in financial risk management owing to its desirable theoretical properties.
However, existing estimation methods for conditional EVaR  either rely on potentially misspecified parametric models or suffer from the curse of dimensionality in nonparametric approaches. 
%However, the existing parametric conditional EVaR estimation may be mis-specified,  and the nonparametric conditional EVaR estimation suffers from the curse of dimensionality. 
To this end, we study semiparametric estimation and inference for conditional EVaR in nonlinear time series models with unspecified error distribution. 
Specifically, 
we propose a two-step conditional EVaR estimator  that achieves the parametric convergence rate. 
Furthermore, 
we derive a functional central limit theorem for conditional EVaR process,  
and then extend to investigate  estimating both conditional expectile-based expected shortfall and conditional distribution. 
A residual bootstrap procedure is proposed for non-expert practitioners to conduct statistical inference.  
The asymptotic results and their applications are illustrated by numerical studies and an empirical application to S\&P 500 returns.

\noindent 
\textbf{Keywords:} 
Conditional EVaR;
Filtered historical simulation;
Nonlinear time series;
Semiparametric methods.

\end{spacing}

\newpage

%%%%%%%%%%%%%%%%%%%%%%%%%%%%%%%%%%%%%%%%%%%%%%%%%%	
\section{Introduction}
\label{s:intro}	

Risk management plays a crucial role in financial institutions and is primarily supported by quantitative risk measurement techniques \citep{engle2004caviar}. 
There are widely used risk measures, including  value-at-risk (VaR), expected shortfall (ES), and expectile-based value-at-risk (EVaR). 
Among of them, 
EVaR is the only law-invariant, elicitable and coherent risk measure \citep{bellini_elicitable_2015, ziegel_coherence_2016, nolde2017elicitability}.
These desirable theoretical properties have stimulated a growing literature on EVaR in both theoretical and empirical settings, 
see 
\citet{kuan_assessing_2009, daouia2021expecthill,  zaevski2023from, zhang2023nonparametric,zhang2025new}, among others. 

In empirical applications, 
a conditional EVaR that adapts to current market conditions is more informative than an unconditional measure \citep{li2023simultaneous, zhang2025new}. 
For instance, market participants, particularly active traders, often rely on short-term signals such as recent asset returns and contemporaneous macroeconomic or global financial indicators when assessing downside risk. 
This motivates the development of econometric methods for estimating and conducting inference on conditional EVaR. 
Most existing conditional EVaR estimation methods are mainly parametric with the specific model form, 
including linear expectile regression models \citep{newey_asymmetric_1987, efron1991regression, taylor_estimating_2008, kuan_assessing_2009}, 
and 
nonlinear expectile regression models \citep{kim_nonlinear_2016, zhang2017continuous, zhang2025threshold}. 
Although these approaches provide interpretability and statistical efficiency under correct specification, 
they are potentially vulnerable to model misspecification due to the imposed functional form. 
To this end, there is a vast amount of literature on nonparametric approaches for estimating conditional EVaR. 
These methods include kernel smoothing estimation %using local smoothing techniques
~\citep{yao1996asymmetric, xie_varying-coefficient_2014, yang2018flexible, %mohammedi2021consistency, 
adam2022local, zhang2025nonparametric}, 
spline-based methods~\citep{de2009quantiles, schnabel2009optimal}, and machine learning method%~\citep{yang2015nonparametric,jiang2017expectile}
~\citep{yang2015nonparametric}.

Despite they can capture nonlinear relationships without restrictive model assumptions, 
nonparametric methods suffer from slow convergence rates, 
sensitivity to tuning parameters and poor scalability in moderate to high dimensions due to the curse of dimensionality \citep{xie_varying-coefficient_2014,adam2022local}. 
To retain the flexibility of nonparametric methods but preserve the structure and efficiency of parametric models, 
several studies consider semiparametric methods for conditional EVaR estimation, 
see~\cite{schulze2017smooth}, %\cite{cai2018assessing}, 
\cite{jiang_single-index_2022}, and among others.
\cite{zhang2025new} proposed a new semiparametric estimator under a linear predictive regression, 
by combining the linear prediction by the least squares estimator and the unconditional EVaR of the least squares residuals. 
However, 
their linearity assumption is too restrictive to adequately capture nonlinear, asymmetric and potentially highly complicated relationships in financial variables 
\citep{cai2012semiparametric,gu2021autoencoder}. 
%Moreover, standard linear models typically assume homoskedastic errors and therefore fail to accommodate volatility clustering, which is one of the most prominent stylized features of financial asset returns \citep{ning2015volatility}. 

In this paper, 
we contribute to the expanding literature on developing semiparametric estimation and inference for conditional EVaR in nonlinear time series models. 
%Our contributions are threefold. 
Our first (and most important) contribution is to propose a novel two-step conditional EVaR estimator, 
which retains the model specification 
and leaves only the error distribution unspecified. 
More specifically, 
it is a combination of the quasi-maximum likelihood estimator for a location-scale model and the empirical EVaR of the filtered residuals.  
%It firstly estimate conditional mean and variance parameters by  quasi-maximum likelihood, and then the conditional expectile is obtained from  the empirical distribution of fitted residuals. 
This idea is essentially the filtered historical simulation (FHS) method, 
which has been widely applied in financial econometrics,  
see for example
conditional VaR estimation~\citep{so2015statistical, francq2015risk, wang2016conditional}, 
options pricing~\citep{barone2008garch}, 
and volatility forecasting~\citep{jiang2022forecasting}.
Notably, our proposed semiparametric estimator has several appealing features.
To start with, 
the parametric model structure allows modeling of return dynamics with covariates, 
and the resulting estimator achieves the parametric $\sqrt{n}$-convergence rate. 
This mitigates the curse of dimensionality and the slow convergence typically associated with nonparametric approaches.
Besides, 
leaving the error distribution unspecified reduces the risk of distributional misspecification.
Moreover, the two-step procedure is easy to implement and sufficiently flexible to accommodate heavy-tailed errors after accounting for conditional heteroskedasticity.

Our second  contribution is to develop   statistical inference of conditional EVaR. 
Specifically, we establish the asymptotic theory and derive consistent standard error estimators for the proposed semiparametric conditional EVaR estimator.
To enable simultaneous inference across multiple expectile levels, we further establish a functional central limit theorem for the conditional EVaR process. 
For practical implementation, we introduce a user-friendly bootstrap procedure for constructing confidence intervals. Simulation results show that the bootstrap confidence intervals exhibit superior finite sample performance relative to the asymptotic confidence intervals.

Last but not least, 
to enhance the practical relevance of our theory, 
the third main contribution is to illustrate the broad applicability of the proposed estimator by two useful extensions.  
One extension concerns semiparametric estimation of conditional expectile-based expected shortfall (XES), 
which was firstly introduced by \citet{daouia2020tailexpectile}. 
Another extension focuses on semiparametric estimation of the conditional distribution, using the one-to-one mapping between the expectile curve and the distribution function \citep{yao1996asymmetric, waltrup2015expectile}. 
For each extension, the corresponding asymptotic properties are  established. 
An empirical study using S\&P~500 returns confirms the great predictive performance of the proposed estimator for conditional EVaR, XES, and distributional forecasts.

The rest of the paper is organized as follows.
Section~\ref{s:est} proposed the semiparametric conditional EVaR estimator and derives its asymptotic properties. 
A residual bootstrap for practical implementation is also provided in this section.  
Section~\ref{s:process} investigates the EVaR process,  
with extensions to  conditional XES and the conditional distribution  recovered from the estimated expectile curves.  
Simulation evidence and an empirical illustration are reported in Sections~\ref{s:sim} and \ref{s:app}, respectively.
Section~\ref{s:con} concludes the paper.
%Proofs of the main results are relegated to the Appendix. 
Proofs of the main results, auxiliary lemmas with their proofs, as well as detailed numerical results, are included in the Supplementary Material.

\section{Methodology}
\label{s:est}

\subsection{Semiparametric conditional  EVaR estimation} 
\label{ss:lse}	
Let $Y_t\in\mathbb{R}$ be the scalar-valued return of an asset or a financial variable, 
and $\bX_t=(X_{t1},\ldots, X_{tp})^\top$ be a $p$-dimensional vector of predictors whose first element is 1. 
In practice, $\bX_t$ may include market covariates or risk factors,  and may also contain lagged values of $Y_t$. 
Assume that the observed sample $\{(Y_t,\bX_t): t=1,\ldots,n\}$ follows the location-scale model, 
\begin{equation}
    \label{eqn:model}
    Y_t= m(\bX_{t}, \bbeta) + \sigma(\bX_{t}, \bbeta)e_t, 
\end{equation}
where $\{e_t\}$ are independent and identically distributed (IID) errors with zero mean and unit variance, 
$m(\cdot, \bbeta)$ and $\sigma(\cdot,\bbeta)>0$ are measurable parametric functions with an unknown $q$-dimensional parameter 
$\bbeta\in\mathbb{R}^q$. 
For the true parameter $\bbeta_0$ in model~\eqref{eqn:model}, 
we have  $\operatorname{E}[Y_t|\bX_t=\bx]=m(\bx, \bbeta_0)$
and $\operatorname{Var}\left(Y_t|\bX_t=\bx\right)=\sigma^2(\bx, \bbeta_0)$. 
Notably, 
the model specification in \eqref{eqn:model} 
is a nonlinear regression model with heteroscedastic errors, 
and is flexible enough to allow $\sigma(\bX_t,\bbeta)$ for taking form of the ARCH and GARCH-type volatility models. 
Therefore, model~\eqref{eqn:model} is widely used in financial empirical work \citep{martins2018nonparametric, girard2021extreme, zhao2026minimax}.  

In this paper, we focus on 
the $\tau$-th conditional EVaR of $Y_t$ given $\bX_t=\bx$,
%which is defined by 
$$
\operatorname{EVaR}_\tau(\bx)
:=\arg\min\limits_y \operatorname{E}\left[\Psi_\tau(Y_t-y)|\bX_t=\bx\right],  
\quad 0<\tau<1, 
$$
where 
$\Psi_\tau(u)=w_\tau(u)u^2$ is 
 the expectile loss 
with $w_\tau(u)=|\tau-I(u\leq 0)|$, 
and $I(\cdot)$ is the indicator function. 
Then, 
under model~\eqref{eqn:model}, 
by the
translation invariance property of the EVaR, 
we have 
\begin{equation}
\label{eqn:evar}
\operatorname{EVaR}_\tau(\bx)
=m(\bX_{t}, \bbeta) + \sigma(\bX_{t}, \bbeta)\mu_\tau(e_t), 
\end{equation}
where  
%the minimizer of an asymmetric quadratic loss 
$
\mu_\tau(e_t)=\arg\min\limits_e \operatorname{E}\left[\Psi_\tau(e_t-e)\right]
$
is the $\tau$th expectile of $e_t$. 

Because $\operatorname{EVaR}_\tau(\bx)$ 
involves the heteroskedastic component $\sigma(\bX_{t}, \bbeta)$,
the conventional linear expectile regression method for estimating $\operatorname{EVaR}_\tau(\bx)$  in \eqref{eqn:evar}  is not directly applicable in this setting. 
Alternatively,   
the nonlinear expectile regression method by  \cite{kim_nonlinear_2016} for jointly estimating $\bbeta$ and $\mu_\tau(e_t)$ is only suitable for a special case of model~\eqref{eqn:model} with $\sigma(\bX_t,\bbeta)\equiv 1$. 
Moreover,  
their optimization procedure is computationally involved and therefore is less attractive in practice.
Therefore,
one of the main research questions addressed in this paper is 
how to estimate $\operatorname{EVaR}_\tau(\bx)$  in \eqref{eqn:evar}. 

Note that asset returns in financial time series 
typically exhibit both a predictable conditional mean structure and pronounced time-varying volatility, 
whereas tail risk is primarily driven by the distributional shape of the innovations. 
Once systematic components in the mean and variance have been filtered out, the remaining uncertainty is mainly reflected in the distribution of the innovation term, allowing tail risk to be estimated more efficiently from the stabilized residual series.  
Therefore,  
by employing the structural decomposition in \eqref{eqn:evar}, 
we propose to estimate the systematic components and the innovation distribution separately. 
Specifically, 
we propose the following two-step estimation procedure: 
\begin{description}
    \item[Step 1:]
    Estimate $\bbeta$ by the quasi maximum likelihood estimator (QMLE): 
    $$
    \widehat{\bbeta}:=
    \arg\min_{\bbeta}\sum_{t=1}^n 
    \left\{ \left[\frac{Y_t-m(\bX_t,\bbeta)}{\sigma(\bX_t,\bbeta)}
    \right]^2 +2 \log \sigma(\bX_t,\bbeta) \right\}.
    $$ 

    \item[Step 2:]
    Estimate $\mu_\tau$ by the $\tau$-th sample expectile $\widehat{\mu}_\tau(\widehat{e}_t)$ of the resulting QMLE residuals $\widehat{e}_t=e_t(\widehat{\bbeta})$, 
    where $e_t(\bbeta)=\left(Y_t-m(\bX_t,\bbeta)\right)/\sigma(\bX_t,\bbeta)$, and 
    \begin{equation}
    \label{eqn:EVaR-e}
    \widehat{\mu}_\tau(\widehat{e}_t)   =\arg\min_{e}n^{-1}\sum_{t=1}^n \Psi_\tau(\widehat{e}_t-e).
    \end{equation}      
    Then,  the proposed  conditional EVaR estimator is given by  
    \begin{equation}
    \label{eqn:est}    \widehat{\operatorname{EVaR}}_\tau(\bx) 
    = m(\bx,\widehat{\bbeta}) +\sigma(\bx,\widehat{\bbeta})\widehat{\mu}_\tau(\widehat{e}_t). 
    \end{equation}
\end{description}

Notably, 
the resulting conditional EVaR estimator in \eqref{eqn:est} is semiparametric. 
It also achieves a trade-off between model structure and model flexibility.
Parametric modeling captures the conditional mean and volatility dynamics, 
whereas the nonparametric expectile estimation applied to filtered residuals avoids imposing restrictive assumptions on the innovation distribution. 
Thus, it can improve robustness in empirical financial applications.   
This two-step approach is commonly referred to as filtered historical simulation (FHS) in the risk management literature~\citep{barone1998market, francq2015risk, so2015statistical, wang2016conditional, francq2020virtual,zhang2025new}. 
More importantly,    
it is not only easy to implement 
but also sufficiently flexible to accommodate a broad class of specifications for $m(\cdot,\bbeta)$ and $\sigma(\cdot,\bbeta)$.

\textbf{Remark 1}. 
When $m(\bX_t,\bbeta)=\bX_t^\top\bbeta$ and $\sigma(\bX_t,\bbeta)=1$,
model~\eqref{eqn:model} reduces to the homoscedastic linear regression model adopted by \cite{zhang2025new}. 
%They consider the ordinary least squares (OLS) estimator in the first step.
However, from a statistical point of view, 
our current extension is far from trivial. 
First, 
by allowing for nonlinear conditional mean structures as well as conditional volatility, 
the proposed conditional EVaR estimator is beyond their linear predictive regression.  
Second, 
compared with their least-squares residual estimator, 
our proposed semiparametric EVaR estimator relies on a reparameterization of the model and the use of a non-Gaussian QMLE. 
This typically entails model flexibility for accommodating 
heavy-tailed errors in financial time series,  
as we will see in more details in numerical studies.  

\subsection{Asymptotic distribution} 
\label{ss:large}	

In this section, we establish the weak convergence of the process $\sqrt{n}[\widehat{\operatorname{EVaR}}_\tau(\bx)-\operatorname{EVaR}_\tau(\bx)]$ indexed by $\tau$,  
under the weakest serial dependence described by $\alpha$-mixing assumption. 
Please refer to \cite{fan2003nonlinear} for more details on $\alpha$-mixing. 
\iffalse
The stationary process $\{(Y_t,\bX_t), t=1,2,\ldots\}$ is called strongly mixing or $\alpha$-mixing \citep{rosenblatt1956central} if 
    $$ 
    \alpha (t)\equiv \sup\limits_{A\in \mathcal{F}_{-\infty }^{0},\,B\in \mathcal{F}%
	_{t}^{\infty }}|P(AB)-P(A)P(B)|\longrightarrow 0, \quad \text{as $t\rightarrow \infty$},
    $$
where $\mathcal{F}_{t_1}^{t_2}$ is the $\sigma $-algebra of events generated by the random variables $\{(\bX_{t}, Y_{t}):t_1\leq t\leq t_2\}$.
\fi          
We firstly impose the following regularity conditions. 
\begin{description}
    \item[(C1)] 
    The process 
    $\{(Y_t, \bX_t): t\geq 1 \}$ 
    is a strictly  stationary $\alpha$-mixing with the mixing coefficients  satisfying $\sum\limits_{t=1}^\infty \alpha(t)^{1-2/\delta}<\infty$ for some $\delta>2$. 
    In addition, $\{e_t\}$ and $\{\bX_t\}$ are independent, 
    and $\{e_t\}$ are independent and identically distributed (IID) errors with zero mean and unit variance. 
    
    \item[(C2)] 
    $\operatorname{E}\|\bX_t\|^{\delta}<\infty$, $\operatorname{E}|Y_t|^{\delta}<\infty$ and $\operatorname{E}|e_t|^{\delta}<\infty$, 
    where $\delta>2$ is given in Condition (C1).

    \item[(C3)]
    Both $m(\cdot,\bbeta)$ and $\sigma(\cdot,\bbeta)>0$ are measurable and continuously differentiable functions with respect to $\bbeta$.

    \item[(C4)] 
    The conditional distribution function $F_{Y_t}(y|\bx)$ of $Y_t$ given $\bX_t=\bx$ is continuous in a neighborhood of 
    $y\in\big[\mu_a(Y|\bx),\,\mu_b(Y|\bx)\big]$ for $\tau\in\mathcal{T}=[a,b]$ with $0<a<b<1$, and admits a continuous density satisfying $0<f_{Y_t}(y|\bx)<\infty$.

    \item[(C5)] 
    For each $\tau\in\mathcal{T}$, 
    $A_\tau(\bbeta)=\operatorname{E}\left[w_\tau(e_t(\bbeta)-\mu_\tau)\right]>0$. 
    The matrix 
    $$
    \ell(\bbeta) = 
    \operatorname{E}\left[\frac{\dot{m}(\bbeta,\bX_1)\dot{m}(\bbeta,\bX_1)^\top
    +2\dot{\sigma}(\bbeta,\bX_1)\dot{\sigma}(\bbeta,\bX_1)^\top}
    {\sigma^2(\bbeta,\bX_1)}\right]
    $$ 
    is positive definite, 
    where $\dot{m}(\bbeta,\bX_1)=\partial m(\bbeta,\bX_1)/\partial \bbeta$ and  $\dot{\sigma}(\bbeta,\bX_1)=\partial\sigma(\bbeta,\bX_1)/\partial \bbeta$.

\end{description}

Conditions (C1)–(C5) are standard for the EVaR estimator in the
time series context.
%see for example %\cite{zhang2025GrangerExpectile} and \cite{zhang2023nonparametric}.
Condition (C1) restricts serial dependence in the time series, which is reasonably weak.
Condition (C2) imposes finite moment requirements that are routine for establishing limiting distributions and stochastic equicontinuity.
Condition (C3) guarantees valid first-order Taylor expansions and enables the Bahadur representation.
Condition (C4) requires the continuity and boundedness of the conditional density function $f_{Y_t}(y|\bx)$. It ensures that the limiting distribution is normal.
Condition (C5) contains standard regularity conditions for expectile estimation that ensure the consistency and asymptotic normality.

% Define $\dot{\Psi}_\tau(u)=uw_\tau(u)$ and $\ddot{\Psi}_\tau(u)=w_\tau(u)$, with $w_\tau(u)=|\tau-I(u\leq 0)|$.  
%Define $\psi_\tau(u)=uw_\tau(u)$ with $w_\tau(u)=|\tau-I(u\leq 0)|$.
The following theorem presents the weak convergence of the proposed semiparametric estimator $\widehat{\operatorname{EVaR}}$. 

\begin{theorem}
\label{thm:asn}	
    Under the Conditions (C1)--(C5), 
    as $n\longrightarrow \infty$, 
    given $\bX_t=\bx$,  
    \begin{description}
        \item[(i)]
        $\widehat{\operatorname{EVaR}}_\tau(\bx)$ is $\sqrt{n}$-consistent, i.e., $\widehat{\operatorname{EVaR}}_\tau(\bx)=\operatorname{EVaR}_\tau(\bx)+O_P(n^{-1/2})$. 

        \item[(ii)]
        $\sqrt{n}\left[\widehat{\operatorname{EVaR}}_\tau(\bx)-
        \operatorname{EVaR}_\tau(\bx)
        \right]$ has the linear representation
	\begin{align}
	\label{eqn:asylinear}
		\sqrt{n}\left[\widehat{\operatorname{EVaR}}_\tau(\bx)-
        \operatorname{EVaR}_\tau(\bx)
        \right]
		=\sigma(\bx,\bbeta_0) \frac{1}{\sqrt{n}} 
             \sum_{t=1}^n Z_t(\tau,\bx, \bbeta_0) 
             + r_n(\tau),
	\end{align}
        where 
        $\sup\limits_{\tau\in \mathcal{T}} |r_n(\tau)|=o_P(1)$, 
        and
        \begin{equation*}
            \begin{aligned}
             Z_t(\tau,\bx,\bbeta)
             ={}&
             A_\tau(\bbeta)^{-1}
             \psi_\tau\left(e_t(\bbeta)-\mu_\tau\right) \\
             &+
             \left[
             \frac{
             \dot{m}(\bx,\bbeta)
             +\dot{\sigma}(\bx,\bbeta)\mu_\tau(e_t(\bbeta))
             }{
             \sigma(\bx,\bbeta)
             }
             +
             A_\tau(\bbeta)^{-1}B_\tau(\bbeta)
             \right]^\top
             D(\bbeta,e_t(\bbeta),\bX_t).
             \end{aligned}
        \end{equation*}
        $\psi_\tau(u)=uw_\tau(u)$, 
        $B_\tau(\bbeta) = \operatorname{E}\left[ w_\tau(e_t(\bbeta)-\mu_\tau) \dot{e}_t(\bbeta)\right]$, 
        $$
        \dot{e}_t(\bbeta)=\frac{-\dot{m}(\bX_t,\bbeta)-e_t(\bbeta)\dot{\sigma}(\bX_t,\bbeta)}{\sigma(\bX_t,\bbeta)},
        $$
        and 
        $$
        D(\bbeta,e_t,\bX_t) =
        \ell(\bbeta)^{-1} \left[\frac{\dot{m}(\bbeta,\bX_t)e_t}{\sigma(\bbeta,\bX_t)}+(e_t^2-1)\frac{\dot{\sigma}(\bbeta,\bX_t)}{\sigma(\bbeta,\bX_t)}\right].
        $$
    \end{description}
     
\end{theorem}

%\noindent \textbf{Remark 1}. 
From Theorem~\ref{thm:asn} (i), the proposed semiparametric EVaR estimator can achieve the parametric convergence rate of $\sqrt{n}$. 
Moreover, $Z_t(\tau,\bx,\bbeta_0)$ in Theorem~\ref{thm:asn} (ii) admits a natural two-part decomposition: 
its first component  captures the variation of the sample expectile of the true innovations $\{e_t\}$, 
whereas its second component  accounts for the estimation uncertainty of the estimator $\widehat{\bbeta}$.

By Theorem~\ref{thm:asn}, 
it is easy to obtain the asymptotic normality of $\widehat{\operatorname{EVaR}}_\tau(\bx)$ that 
\begin{align}
\label{eqn:asn}
\sqrt{n}\left[\widehat{\operatorname{EVaR}}_\tau(\bx)-      \operatorname{EVaR}_\tau(\bx)\right]
{\stackrel{D}{\longrightarrow}}  
N\left(0, \sigma^2(\bx,\bbeta_0) \Gamma(\tau,\bx,\bbeta_0)
\right),
\end{align}
where 
 \begin{equation*}
   \begin{aligned}
    \Gamma(\tau,\bx,\bbeta)
         &=
         \lim_{n\rightarrow\infty} \operatorname{Var}
         \left( \frac{1}{\sqrt{n}} \sum_{t=1}^n Z_t(\tau,\bx, \bbeta) \right)\\
         & =
   \operatorname{E}\left[Z_1^2(\tau,\bx,\bbeta)\right]
   +2\sum\limits_{l=1}^\infty\operatorname{E}\left[Z_1(\tau,\bx,\bbeta)
   Z_{1+l}(\tau,\bx,\bbeta)\right].
 \end{aligned}
\end{equation*}

%\subsection{Consistent estimate of the limiting variance} \label{ss:cov-est}

To make statistical inference for the proposed estimator $\widehat{\operatorname{EVaR}}_\tau(\bx)$, 
%in practice,  
it needs to estimate the asymptotic variance $\sigma^2(\bx,\bbeta_0) \Gamma(\tau,\bx,\bbeta_0)$ in (\ref{eqn:asn}). 
Specifically, 
$\sigma^2(\bx,\bbeta_0)$ can be estimated by 
$\sigma^2(\bx,\widehat{\bbeta})$, 
since $\sigma^2(\bx,\bbeta)$ is continuous with respect to $\bbeta$ in Condition (C3). 
Following 
\citet{de2000consistency}, 
$\Gamma(\tau,\bx,\bbeta_0)$
can be consistently estimated by a kernel heteroskedasticity and autocorrelation
consistent (HAC) form estimator,

\begin{equation*}
    \begin{aligned}
        \widehat{\Gamma}(\tau,\bx,\widehat{\bbeta})
        &=
        \frac{1}{n}\sum_{s=-n+1}^{n-1}\sum_{t=1}^{n-s}
        \widehat Z_{t}(\tau,\bx,\widehat{\bbeta})
        \widehat Z_{t+s}(\tau,\bx,\widehat{\bbeta})K\left(\frac{s}{b_n}\right),
    \end{aligned}
\end{equation*}
where 
\begin{equation*}
\begin{aligned}
\widehat Z_t(\tau,\bx,\widehat{\bbeta})
={}&
\widehat A_\tau(\widehat{\bbeta})^{-1}
\psi_\tau\left(\widehat e_t-\widehat\mu_\tau\right)
\\
&+
\left[
\frac{
\dot m(\bx,\widehat{\bbeta})
+
\dot\sigma(\bx,\widehat{\bbeta})\widehat\mu_\tau
}{
\sigma(\bx,\widehat{\bbeta})
}
+
\widehat A_\tau(\widehat{\bbeta})^{-1}
\widehat B_\tau(\widehat{\bbeta})
\right]^\top
\widehat D(\widehat{\bbeta},e_t(\widehat{\bbeta}),\bX_t),
\end{aligned}
\end{equation*}
with 
$
\widehat A_\tau(\widehat{\bbeta})
=
n^{-1}\sum\limits_{t=1}^n
w_\tau\left(\widehat e_t(\widehat{\bbeta})-\widehat\mu_\tau\right)
$,
$
\widehat B_\tau(\widehat{\bbeta})
=
n^{-1}\sum\limits_{t=1}^n
w_\tau\left(\widehat e_t(\widehat{\bbeta})-\widehat\mu_\tau\right)
\dot e_t(\widehat{\bbeta})
$,
and 
$$
\widehat D(\widehat{\bbeta},e_t(\widehat{\bbeta}),\bX_t)
=
\widehat{\ell}(\widehat{\bbeta})^{-1}
\left[
\frac{
\dot m(\widehat{\bbeta},\bX_t)\widehat e_t
}{
\sigma(\widehat{\bbeta},\bX_t)
}
+
\frac{
(\widehat e_t^2-1)
\dot\sigma(\widehat{\bbeta},\bX_t)
}{
\sigma(\widehat{\bbeta},\bX_t)
}
\right],
$$
$$
\widehat{\ell}(\widehat{\bbeta})
=
n^{-1}\sum\limits_{t=1}^n
\frac{
\dot m(\widehat{\bbeta},\bX_t)
\dot m(\widehat{\bbeta},\bX_t)^\top
+
2\dot\sigma(\widehat{\bbeta},\bX_t)
\dot\sigma(\widehat{\bbeta},\bX_t)^\top
}{
\sigma^2(\widehat{\bbeta},\bX_t)
}.
$$
Here,
$K(\cdot)$ is a kernel function with a lag truncation parameter $b_n>0$.  
To establish the consistency of $\widehat{\Gamma}(\tau,\bx,\widehat{\bbeta})$, 
we need the following condition on kernel and bandwidth. 
\begin{description}
    \item[(C6)]  
    The kernel function $K(\cdot)$ is symmetric with a compact support satisfying $\int_{-\infty}^{\infty} K(u)du=1$, and has a bounded first derivative. The lag truncation parameter satisfies that $\lim\limits_{n\rightarrow\infty} (b_n^{-1}+n^{-1/2}b_n)=0$. 
\end{description}

The following result provides the uniform consistency of $\widehat{\Gamma}(\tau,\bx,\widehat{\bbeta})$ over $\tau$. 

\begin{theorem}
\label{thm:hac}	
    Under the Conditions (C1)--(C6), as $n\longrightarrow \infty$, 
    given $\bX_t=\bx$,  
    $\widehat{\Gamma}(\tau,\bx,\widehat{\bbeta})
    {\stackrel{P}{\longrightarrow}} \Gamma(\tau,\bx,\bbeta_0)$ 
    uniformly over $\tau\in\mathcal{T}$. 
    %where $\mathcal{T}\subset(0,1)$ is a compact index set.      
\end{theorem}

Theorem~\ref{thm:hac} 
guarantees that the proposed HAC estimator consistently estimates the long-run variance of the conditional EVaR estimator under weak serial dependence. 
This result ensures that the confidence intervals remain asymptotically valid in the presence of time-series dependence and conditional heteroskedasticity. 
In our later applications, as suggested in \citet{andrews1991heteroskedasticity}, 
we use the Bartlett kernel $K(u)=(1-|u|)I(|u|\leq 1)$ with $b_n=[4(n/100)^{1/3}]$, where $[a]$ is the integer part of $a$. 
Then, the $(1-\alpha)$-th Wald confidence interval for $\operatorname{EVaR}_\tau(\bx)$ is given by 
\begin{equation}
\label{eq:ci}
\widehat{\operatorname{EVaR}}_\tau(\bx) \;\pm\; z_{\frac{\alpha}{2}} 
\sigma(\bx,\widehat{\bbeta})
n^{-1/2} \widehat{\Gamma}(\tau,\bx,\widehat{\bbeta}) ^{1/2}, 
\end{equation}
where $z_{\alpha}$ is the $\alpha$-th tail percentile of the standard normal distribution $N(0,1)$.

\subsection{Practical implementation} 
\label{ss:implement}

Statistical inference for $\widehat{\operatorname{EVaR}}_\tau(\bx)$ 
requires knowledge of the Bahadur representation of $\widehat{\bbeta}$ as well as a consistent estimator of the long-run covariance matrix $\widehat{\Gamma}(\tau_1,\tau_2)$. 
However, it may be non-trivial for a non-expert practitioner to derive a Bahadur representation for their specific model and parameter estimation. 
To facilitate implementation, this subsection introduces a user-friendly residual bootstrap procedure that avoids explicit analytical derivations~\citep{cavaliere2022bootstrap, beutner2024residual}.
More specifically, we resample the fitted residuals and then generate bootstrap observations while keeping the dynamics of the bootstrap samples fixed at the value of the original series.
The proposed fixed-design residual bootstrap procedure is summarized in Algorithm~\ref{alg:fixed_boot}.
%Assume that the covariates $\bX_t=(Y_{t-1},\ldots, Y_{t-k}, \bZ_t)$ consists of lagged dependent variables and additional exogenous predictors $\bZ_t$, the proposed residual bootstrap procedure is summarised in Algorithm \ref{alg:bootstrap}.
% We propose the following residual bootstrap procedure.

\begin{algorithm}
\caption{Fixed-design residual bootstrap procedure.}
\label{alg:fixed_boot}

\begin{algorithmic}[1]
\setstretch{1.5}

\State
For the original data $\{(Y_t,\bX_t)\}_{t=1}^n$, apply the proposed two-step estimation procedure to obtain $\widehat{\bbeta}$ and the residuals $\widehat e_t$, and then calculate the conditional EVaR estimator $\widehat{\operatorname{EVaR}}_\tau(\bx)$.

\State
Independently draw $\{e_t^*\}$ with replacement from the resulting QMLE residuals residuals $\{\widehat{e}_t\}$, where 
$\widehat{e}_t
=\left(Y_t-m(\bX_t,\widehat{\bbeta})\right)/
\sigma(\bX_t,\widehat{\bbeta})$.

\State
Generate the bootstrap responses while keeping the original design fixed:
\begin{equation}
\label{eq:lm_fixed_boot}
Y_t^*
=
m(\bX_t,\widehat{\bbeta})
+
\sigma(\bX_t,\widehat{\bbeta})e_t^*,
\qquad t=1,\ldots,n.
\end{equation}
Using the fixed-design bootstrap sample $\{(Y_t^*,\bX_t)\}_{t=1}^n$, 
apply the proposed two-step procedure to obtain the bootstrap estimates $\widehat{\bbeta}^*$, $\widehat e_t^*$, and $\widehat{\operatorname{EVaR}}_\tau^*(\bx)$.

\State
Repeat Steps 2 and 3 $M$ times to obtain the bootstrap replications
$
\widehat{\operatorname{EVaR}}_\tau^{*(1)}(\bx),
\ldots,
\widehat{\operatorname{EVaR}}_\tau^{*(M)}(\bx).
$
\end{algorithmic}
\end{algorithm}

\iffalse
\begin{description}
    \item[(i)] 
    Apply the proposed two-step estimation procedure to the original data to obtain $\widehat{\bbeta}$, residuals $\widehat{e}_t$, and the conditional EVaR estimator $\widehat{\operatorname{EVaR}}_\tau(\bx)$.

    \item[(ii)]
    Generate a bootstrap sample $\{(Y_t^*, \bX_t^*)\}_{t=1}^n$ recursively using the fitted model:
    \begin{equation}
    \label{eq:lm_boot}
    Y_t^* = m(\bX_t^*, \widehat{\bbeta}) + \sigma(\bX_t^*, \widehat{\bbeta}) e_t^*, 
    \quad\text{with}\quad \bX_t^* = (Y_{t-1}^*, \ldots, Y_{t-k}^*, \bZ_t),
    \end{equation}
    where $\{e_t^*\}$ are independently drawn with replacement from the empirical residuals $\{\widehat{e}_t\}$. 
    Using the bootstrap sample, reapply the two-step procedure to obtain bootstrap estimates $\widehat{\bbeta}^*$, $\widehat{e}_t^*$, and $\widehat{\operatorname{EVaR}}_\tau^*(\bx)$.
    
    \item[(iii)]
    Repeat step (ii) $M$ times to obtain bootstrap replications of $\widehat{\operatorname{EVaR}}_\tau^*(\bx)$, denoted by
    $\widehat{\operatorname{EVaR}}_\tau^{*(1)}(\bx),\ldots, \widehat{\operatorname{EVaR}}_\tau^{*(M)}(\bx)$. 
\end{description}
\fi

Then, a bootstrap Wald confidence interval can be naturally constructed using the standard error $\widehat{\mathrm{SE}}_{\mathrm{boot}}(\tau,\bx)$ of 
$M$ (e.g., $M=1000$) bootstrap replications 
$\widehat{\operatorname{EVaR}}_\tau^{*(1)}(\bx),\ldots, \widehat{\operatorname{EVaR}}_\tau^{*(M)}(\bx)$. 
\iffalse
\[
\widehat{\mathrm{SE}}_{\mathrm{boot}}(\tau,\bx)
=
\left(
\frac{1}{M-1}
\sum_{m=1}^M
\left[
\widehat{\operatorname{EVaR}}_\tau^{*(m)}(\bx)
-
\overline{\widehat{\operatorname{EVaR}}_\tau^*(\bx)}
\right]^2
\right)^{1/2},
\]
where 
$
\overline{\widehat{\operatorname{EVaR}}_\tau^*(\bx)}
=
\frac{1}{M}
\sum_{m=1}^M
\widehat{\operatorname{EVaR}}_\tau^{*(m)}(\bx).
$
\fi
Specifically, 
the corresponding $(1-\alpha)$-th bootstrap Wald confidence interval is given by 
\begin{equation}
\label{eq:ci_boot}
    \widehat{\operatorname{EVaR}}_\tau(\bx)
    \pm
    z_{\frac{\alpha}{2}}
    \widehat{\mathrm{SE}}_{\mathrm{boot}}(\tau,\bx).
\end{equation}
%Since  model~\eqref{eq:lm_fixed_boot} preserves the structural form of the original model in \eqref{eqn:model}, the bootstrap samples are able to closely replicate the dependence dynamics present in the observed data. 
The simulation results in Section~\ref{ss:cp} 
show  that the bootstrap confidence intervals \eqref{eq:ci_boot} exhibit superior finite sample performance relative to the asymptotic confidence intervals \eqref{eq:ci}, which rely on estimated long-run variances.

\section{EVaR Process and Two Useful Extensions} 
\label{s:process}

In risk management, practitioners often do not focus on a single expectile level. Rather, they require simultaneous inference over a range of levels to assess tail risk and guide dynamic portfolio adjustments \citep{wang2016conditional}. 
Therefore, 
it is necessary to study the EVaR process, i.e., 
the collection $\{\operatorname{EVaR}_\tau(\bx): \tau\in\mathcal{T}\}$, 
where $\mathcal{T}$ is a compact subset of $(0,1)$. 
The following theorem establishes a functional central limit theorem (CLT) version of Theorem~\ref{thm:asn}.

\begin{theorem}
\label{thm:weak}	
    Under the Conditions (C1)--(C5), 
    as $n\rightarrow \infty$, 
    given $\bX_t=\bx$,  
    we have that 
    $\sqrt{n}\left[\widehat{\operatorname{EVaR}}_\tau(\bx)-        \operatorname{EVaR}_\tau(\bx) \right]$  converges weakly in $C(\mathcal{T})$ to a centered Gaussian process $\sigma(\bx,\bbeta_0)G(\tau, \bx, \bbeta_0)$, where $G(\tau,\bx,\bbeta)$ is a zero-mean Gaussian process on $\mathcal{T}$ with continuous sample paths and long-run covariance function
\begin{equation*}
\begin{aligned}
        \Sigma(\tau_1,\tau_2;\bx,\bbeta)
	    = &
		\operatorname{E}\left[Z_t(\tau_1,\bx,\bbeta)Z_t(\tau_2,\bx,\bbeta)\right]
		+
		\sum\limits_{l=1}^\infty
		\operatorname{E}\left[
		Z_t(\tau_1,\bx,\bbeta)
		Z_{t+l}(\tau_2,\bx, \bbeta)
		\right]\\
        &    +
		\sum\limits_{l=1}^\infty
		\operatorname{E}\left[
		Z_{t+l}(\tau_1,\bx,\bbeta)Z_t(\tau_2,\bx,\bbeta)
		\right].
\end{aligned}
\end{equation*}
     
\end{theorem}

%\noindent \textbf{Remark 3.} 
Basically, 
Theorem~\ref{thm:weak} 
extends the pointwise asymptotic normality in Theorem~\ref{thm:asn} to a functional CLT for the entire conditional EVaR curve. 
It enables uniform inference over a continuum of expectile levels and provides a theoretical foundation for constructing simultaneous confidence intervals. 
More importantly,  
the weak convergence for the conditional EVaR process in Theorem~\ref{thm:weak} enables us to consider two useful extensions in what follows.  

\subsection{Conditional expectile-based expected shortfall}
\label{ss:cxes}
We firstly consider an extension to the semiparametric estimation of conditional expectile-based expected shortfall (XES), which was introduced by \citet{daouia2020tailexpectile} and \citet{zhang2023nonparametric}. 
Conditional on $\bX_t=\bx$, the conditional XES is defined as
$$
\operatorname{XES}_\tau(\bx) =
\frac{1}{\tau}\int_0^\tau \operatorname{EVaR}_s(\bx)ds. 
$$
Thus, a natural estimator of $\operatorname{XES}_\tau(\bx)$ is its sample analogue
$$
\widehat{\operatorname{XES}}_\tau(\bx) =
\frac{1}{\tau}\int_0^\tau \widehat{\operatorname{EVaR}}_s(\bx)ds. 
$$

\begin{theorem}
\label{thm:weak_xes}	
    Under the Conditions (C1)--(C5), as $n\longrightarrow \infty$, 
    given $\bX_t=\bx$,     
    we have that 
    $\sqrt{n}
    \left[\widehat{\operatorname{XES}}_\tau(\bx)
    -\operatorname{XES}_\tau(\bx) \right]$ 
    admits the asymptotic linearization
    \begin{align}
    \label{eqn:asylinear_xes}
	\sqrt{n}
    \left[\widehat{\operatorname{XES}}_\tau(\bx)-             
        \operatorname{XES}_\tau(\bx) \right]
	=
        \sigma(\bx,\bbeta_0) \frac{1}{\sqrt{n}} 
        \sum_{t=1}^n \frac{1}{\tau}\int_0^\tau 
        Z_t(s,\bx,\bbeta_0)ds + r_n(\tau), 
    \end{align}
     where $\sup\limits_{\tau\in \mathcal{T}} |r_n(\tau)|=o_P(1)$, 
     and $Z_t(\tau,\bx, \bbeta)$ is defined in Theorem~\ref{thm:asn}.  
     Furthermore, 
     $\sqrt{n}
    \left[\widehat{\operatorname{XES}}_\tau(\bx)
    -\operatorname{XES}_\tau(\bx) \right]$ 
    converges weakly to a centered Gaussian process $\sigma(\bx,\bbeta_0)\tau^{-1}\int_0^\tau G(s,\bx,\bbeta_0)ds$, 
    where $G(\cdot)$ is the Gaussian process in Theorem~\ref{thm:weak}.
\end{theorem}

%\noindent \textbf{Remark 4.} 
The asymptotic behavior of the proposed XES estimator is directly inherited from that of the EVaR process through a linear functional mapping. 
As a result, statistical inference for conditional XES can be conducted using the same Gaussian limit process without requiring additional estimation of specific tail quantities. 
Moreover, it has been shown that mappings exist from expectiles to quantiles 
\citep{efron1991regression, yao1996asymmetric} and 
quantile-based  expected shortfall   \citep{taylor_estimating_2008,taylor2022forecasting}. Therefore, one can also use our proposed conditional EVaR estimator to estimate the quantile-based risk measures, VaR and ES.

\subsection{Conditional distribution estimation}
\label{ss:cde}

The distribution of financial returns is a relevant object for making financial decisions, as it provides a complete characterization of potential future outcomes \citep{hua2013forecasting}.
Motivated by the fact that expectiles can uniquely define a distribution function \citep{taylor_estimating_2008,waltrup2015expectile}, 
we consider another extension in this subsection to recover the conditional distribution of $Y_t$ given $\bX_t=\bx$ directly from the estimated conditional EVaR curve
$\{\operatorname{EVaR}_\tau(\bx):\tau\in(0,1)\}$.

As showed in \citet{yao1996asymmetric}, $\operatorname{EVaR}_\tau(\bx)$ is linked to the conditional distribution
$F(\cdot|\bx)$ through
\begin{equation}
\label{eq:cond-expectile-dist}
%\resizebox{0.9\textwidth}{!}{$
\operatorname{EVaR}_\tau(\bx)
=
\frac{(1-\tau)\,M\left(\operatorname{EVaR}_\tau(\bx)|\bx\right)
+\tau\left[\operatorname{EVaR}_{0.5}(\bx)-M\left(\operatorname{EVaR}_\tau(\bx)|\bx\right)\right]}
{(1-\tau)\,F\left(\operatorname{EVaR}_\tau(\bx)|\bx\right)
+\tau\left[1-F\left(\operatorname{EVaR}_\tau(\bx)|\bx\right)\right]},
%$}
\end{equation}
where $M(y|\bx)=\int_{-\infty}^{y} z\,\mathrm{d}F(z|\bx)$ is the partial moment function.  
This relationship in \eqref{eq:cond-expectile-dist} provides a one-to-one mapping between the entire expectile curve $\{\operatorname{EVaR}_\tau(\bx):\tau\in(0,1)\}$ and the conditional distribution $F(\cdot|\bx)$.
Although solving \eqref{eq:cond-expectile-dist} for $F(\cdot|\bx)$ is not available in closed form, it can be carried out numerically.

To facilitate implementation, 
we  propose an estimator of $F(\cdot|\bx)$ by $\operatorname{EVaR}_\tau(\bx)$, denoted by  
$\widehat{F}(\cdot|\bx)$.  
Motivated by \citet{waltrup2015expectile}, 
a simple computational procedure~\footnote{One can be recommended to use the R package~\textit{expectreg} by \cite{sobotka2022expectreg} for practical implementation.} 
is summarized in Algorithm~\ref{alg:dist}.

\begin{algorithm}
\caption{Estimate $\widehat F(\cdot|\bx)$ by $\operatorname{EVaR}_\tau(\bx)$.}
\label{alg:dist}

\begin{algorithmic}[1]
\setstretch{1.5}
\State  
Estimate $\widehat{\operatorname{EVaR}}_l(\bx)
= \widehat{\operatorname{EVaR}}_{\tau_l}(\bx)$
on a dense grid $\{\tau_l\}_{l=1}^{L} \subset (0,1)$.

\State
Estimate 
$\widehat F\left(\widehat{\operatorname{EVaR}}_l(\bx)|\bx\right)
=\sum\limits_{j=1}^{l}\widehat\delta_j(\bx)$, 
where  
$\{\widehat\delta_l(\bx)\}_{l=1}^{L}$ 
is determined by solving the system of nonlinear equations:
\begin{equation*}
\widehat{\operatorname{EVaR}}_l(\bx)
=
\frac{
(1-\tau_l)\sum_{j=1}^{l}\widehat d_j(\bx)\widehat\delta_j(\bx)
+\tau_l\left[\widehat{\operatorname{EVaR}}_{0.5}(\bx)-\sum_{j=1}^{l}\widehat d_j(\bx)\widehat\delta_j(\bx)\right]
}
{
(1-\tau_l)\sum_{j=1}^{l}\widehat\delta_j(\bx)
+
\tau_l\left[1-\sum_{j=1}^{l}\widehat\delta_j(\bx)\right]},
\end{equation*}
and $\widehat d_j(\bx)=
0.5[\widehat{\operatorname{EVaR}}_j(\bx)-\widehat{\operatorname{EVaR}}_{j-1}(\bx)]$. 
Set 
$\widehat{\operatorname{EVaR}}_{L+1}(\bx)=\widehat{\operatorname{EVaR}}_{0.5}(\bx)$ for identification.

\State
Construct $\widehat F(\cdot|\bx)$ by linear interpolation between adjacent support points.

\end{algorithmic}
\end{algorithm}

\iffalse
\begin{description}
\item [(i)]
Estimate $\widehat{\operatorname{EVaR}}_l(\bx)
= \widehat{\operatorname{EVaR}}_{\tau_l}(\bx)$
on a dense grid $\{\tau_l\}_{l=1}^L \subset (0,1)$.

\item [(ii)]
Approximate $\widehat F(\cdot|\bx)$ by a step function with nonnegative increments
$\{\widehat\delta_l(\bx)\}_{l=1}^{L+1}$ satisfying $\sum_{l=1}^{L+1}\widehat\delta_l(\bx)=1$.
The fitted conditional distribution function evaluated at the support points is then given by
$\widehat F\big(\widehat{\operatorname{EVaR}}_l|\bx\big)
=\sum_{j=1}^{l}\widehat\delta_j(\bx)$.

\item [(iii)]
Determine $\{\widehat\delta_l(\bx)\}_{l=1}^{L}$ by solving the system of nonlinear equations:
\begin{equation*}
\widehat{\operatorname{EVaR}}_l(\bx)
=
\frac{(1-\tau_l)\sum_{j=1}^{l}\widehat d_j(\bx)\widehat\delta_j(\bx)
+\tau_l\Big(\widehat{\operatorname{EVaR}}_{0.5}(\bx)-\sum_{j=1}^{l}\widehat d_j(\bx)\widehat\delta_j(\bx)\Big)}
{(1-\tau_l)\sum_{j=1}^{l}\widehat\delta_j(\bx)
+\tau_l\Big(1-\sum_{j=1}^{l}\widehat\delta_j(\bx)\Big)},
\end{equation*}
where $\widehat d_j(\bx)=\{\widehat{\operatorname{EVaR}}_j(\bx)-\widehat{\operatorname{EVaR}}_{j-1}(\bx)\}/2$ 
with the identifying constraint 
$\widehat M_{L+1}(\bx)=\widehat{\operatorname{EVaR}}_{0.5}(\bx)$.

\item [(iv)]
Construct $\widehat F(\cdot|\bx)$ by linear interpolation between adjacent support points.
\end{description}
\fi

To establish the asymptotic properties of the resulting estimator $\widehat F(\cdot|\bx)$, we impose the following additional condition.

\begin{description}
    \item[(C7)] 
    The conditional distribution $F(\cdot|\bx)$ admits a density $f(\cdot|\bx)$ that is continuous and satisfies
    $0<\underline f\le f(y\mid\bx)\le \bar f<\infty$ for all $y$ in an open neighborhood of $\mathcal{Y}:=\left\{
    \operatorname{EVaR}_\tau(\bx):\tau\in \mathcal{T}\right\}$,
    where $\mathcal{T}$ is given in Condition (C4). 
\end{description}

To proceed, 
given $\tau\in(0,1)$ and $\bX_t=\bx$,  
for any conditional distribution $F(\cdot|\bx)$ and 
any risk measure $\nu_\tau(\bx)$, 
we define the functional
\begin{equation}
\label{eq:Phi_def}
\Phi 
\left(F(\cdot|\bx), \nu_\tau(\bx)\right)
:=
\nu_\tau(\bx)
-
\frac{ (1-\tau)M(\nu_\tau(\bx)|\bx)
+
\tau\left[\nu_{0.5}(\bx)-M(\nu_\tau(\bx)|\bx)\right]
}{
(1-\tau)F(\nu_\tau(\bx)|\bx)
+
\tau\left[1-F(\nu_\tau(\bx)|\bx)\right]
}.
\end{equation}
Clearly, 
$\Phi\big
\{F(\cdot|\bx),\operatorname{EVaR}_\tau(\bx)\big\}
=0$.

The following theorem establishes the weak convergence of $\widehat F(\cdot|\bx)$.
\begin{theorem}
\label{thm:weak_F}
Under Conditions (C1)--(C5) and (C7),  as $n\to\infty$, 
given $\bX_t=\bx$, 
we have that 
$\sqrt{n}\left[\widehat F(\cdot|\bx)- F(\cdot|\bx) \right]$  converges weakly in $\ell^\infty(\mathcal{Y})$ to a centered Gaussian process $\dot{\mathcal{S}}_{\operatorname{EVaR}_\tau(\bx)}\left[\sigma(\bx,\bbeta_0)\,G(\tau, \bx,\bbeta_0)\right]$, 
where 
the solution mapping 
$\mathcal{S}: C(\mathcal{T})\to\ell^\infty(\mathcal{Y})$ is  implicitly defined by $\Phi\big
\{F(\cdot|\bx),\operatorname{EVaR}_\tau(\bx)\big\}
=0$, 
and $G(\cdot)$ is the Gaussian process defined 
in Theorem~\ref{thm:weak}. 
Moreover, $\mathcal{S}$ is Hadamard differentiable at $\operatorname{EVaR}_\tau(\bx)$, 
with derivative
$\dot{\mathcal{S}}_{\operatorname{EVaR}_\tau(\bx)}[h]\in\ell^\infty(\mathcal{Y})$
characterized by
$\dot\Phi_{F,\operatorname{EVaR}_\tau(\bx)}\big[\dot{\mathcal{S}}_{\operatorname{EVaR}_\tau(\bx)}[h],\,h\big]=0$
for every $h\in C(\mathcal{T})$,
where the Hadamard derivative $\dot\Phi_{F,\operatorname{EVaR}_\tau(\bx)}$ admits the explicit expansion
\begin{align}
\label{eq:dotPhi}
&~~~~
\dot\Phi_{F,\operatorname{EVaR}_\tau(\bx)}[H,h]\nonumber\\
&=
h(\tau)
-
\frac{
\tau\,h(0.5)
+
(1-2\tau)\left[
\int_{-\infty}^{\operatorname{EVaR}_\tau(\bx)} z\,\mathrm{d}H(z|\bx)
-
\operatorname{EVaR}_\tau(\bx)H(\operatorname{EVaR}_\tau(\bx)|\bx)
\right]
}{
(1-\tau)F(\operatorname{EVaR}_\tau(\bx)|\bx)+\tau\{1-F(\operatorname{EVaR}_\tau(\bx)|\bx)\}
},
\end{align}
for any directions $H(\cdot|\bx)\in\ell^\infty(\mathbb{R})$ and $h\in C(\mathcal{T})$.
\end{theorem}

%\noindent\textbf{Remark 5.}
Theorem~\ref{thm:weak_F} 
shows that the asymptotic behavior of $\widehat F(\cdot|\bx)$ is entirely driven by that of
the EVaR process through the solution mapping $\mathcal{S}$. 
It entails several attractive features. 
First, the explicit form of the Hadamard derivative in \eqref{eq:dotPhi} highlights how local perturbations of the EVaR curve over $\tau\in \mathcal{T}$ propagate to the estimated conditional distribution function via a tail-integral operator.
It reflects the fact that expectiles are global tail characteristics 
rather than pointwise features of the distribution.
Second, Theorem~\ref{thm:weak_F} implies the asymptotic normality of $\widehat F(\cdot|\bx)$.
Building on this result, 
the Gaussian process limit provides a convenient basis for statistical inference on $\widehat F(\cdot|\bx)$, including the construction of confidence intervals.

\section{Numerical Studies} 
\label{s:sim} 

In this section, we conduct several simulation studies to illustrate the good finite-sample performance of our proposed estimator. 
We first investigate  numerical comparisons of our proposed semiparametric estimator with nonparametric estimators and parametric estimators. 
Next, 
we consider the performance of the proposed semiparametric EVaR estimator under model misspecification, 
and assessing the validity of the bootstrap confidence intervals.

\subsection{Comparison with nonparametric estimators}
\label{ss:comparison}

To assess the performance of the proposed semiparametric EVaR estimator 
across different parametric models, 
we consider four increasingly complex data generating processes (DGPs).
\begin{itemize}
   \item[DGP~1] 
    AR(1)-ARCH(1) model: 
    \(
        Y_{t}=\beta_0+ \beta_1Y_{t-1} +\sqrt{\beta_2^2+\beta_3^2 Y_{t-1}^2} \varepsilon_t, 
    \)   
   where $(\beta_0,\beta_1,\beta_2,\beta_3)=(0.2, 0.7, 0.3, 0.5)$.

   \item[DGP~2] 
   AR(1)-ARCH(1)-X model: 
    \(
        Y_{t}=\beta_0+ \beta_1Y_{t-1} +\gamma_1U_t+\sqrt{\beta_2^2+\beta_3^2 Y_{t-1}^2+\gamma_2^2U_t^2} \varepsilon_t, 
    \)
   where $U_t \sim \mathcal{U}[0,1]$ and $(\beta_0,\beta_1,\beta_2,\beta_3,\gamma_1,\gamma_2)=(0.2,0.7,0.3, 0.5,-0.5,0.5)$.

   \item[DGP~3] 
   GARCH(1,1) model: 
    \(
        Y_{t}= \mu + \sigma_t \varepsilon_t, ~~~
        \sigma_t^2 = \omega + \alpha (Y_{t-1}-\mu)^2 + \beta\sigma_{t-1}^2,
    \)
   where $(\mu,\omega,\alpha,\beta)=(0.5,0.2,0.7,0.1)$.

   \item[DGP~4] 
   GJR-GARCH(1,1) model: 
    \(
         Y_{t}= \mu + \sigma_t \varepsilon_t, 
        \sigma_t^2 = \omega + \alpha (Y_{t-1}-\mu)^2 + \beta\sigma_{t-1}^2 + \gamma (Y_{t-1}-\mu)^2I[(Y_{t-1}-\mu)<0],
    \)
   where $(\mu,\omega,\alpha,\beta,\gamma)=(0.5,0.2,0.7,0.1,0.1)$.

\end{itemize}
The innovation $\varepsilon_t$ is generated from one of the following two settings: 
\begin{description}
    \item[(1)] standard normal $N(0,1)$,
   
    \item[(2)] $t_4/\sqrt{2}$ (Student-$t$ distribution with 4 degrees of freedom with the normalizer $\sqrt{2}$ making the variance one).    
\end{description}

We consider different sample sizes $n\in \{500,1000\}$ and EVaR levels $\tau \in \{0.05, 0.10, 0.25, \allowbreak 0.50, 0.75, 0.90, 0.95 \}$.
All recursive processes are initialized at $Y_0=0$.
For DGPs~3 and 4, the initial conditional variance is set to be its unconditional expectation: 
$\sigma_0^2=\omega/(1-\alpha-\beta)$ for the GARCH(1,1) model and $\sigma_0^2=\omega/(1-\alpha-\beta-\gamma/2)$ for the GJR-GARCH(1,1) model, respectively.
We estimate the $\tau$-th conditional EVaR of $Y_t$ given 
$\bX_t=(1,Y_{t-1})^\top$ and $\bX_t=(1,Y_{t-1},U_{t})^\top$ for DGP~1 and DGP~2, respectively, 
whereas  $\bX_t=(1,Y_{t-1},Y_{t-2},\ldots)^\top$ for DGPs 3 and 4. 
The predictor vector is set as $\bx=(1,0.5)^\top$ for DGP 1 and at $\bx=(1,0.5,0.5)^\top$ for DGP 2,
while 
is set at a prespecified sequence of lagged returns for 
both DGPs~3 and 4. 

\begin{figure}%[htbp]
\centering
\setlength{\tabcolsep}{3pt}
\renewcommand{\arraystretch}{1}
\begin{tabular}{
@{}
c
ccc
@{}
}

% DGP 1
\adjustbox{valign=m}{\rotatebox{90}{\textbf{DGP1}}}
&
\adjustbox{valign=m}{%
  \includegraphics[width=0.31\textwidth]
  {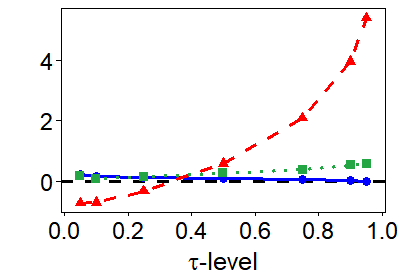}}
&
\adjustbox{valign=m}{%
  \includegraphics[width=0.31\textwidth]
  {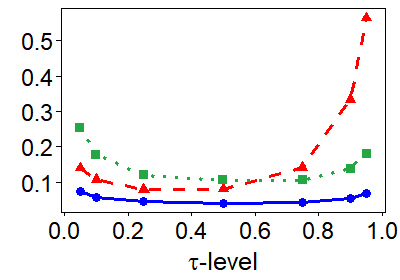}}
&
\adjustbox{valign=m}{%
  \includegraphics[width=0.31\textwidth]
  {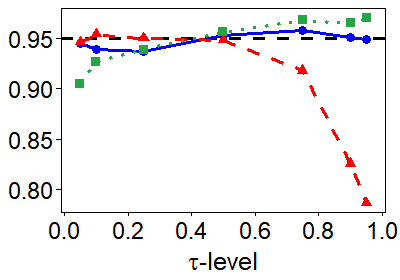}}
\\
\noalign{\vskip 2mm}

% DGP 2
\adjustbox{valign=m}{\rotatebox{90}{\textbf{DGP2}}}
&
\adjustbox{valign=m}{%
  \includegraphics[width=0.31\textwidth]
  {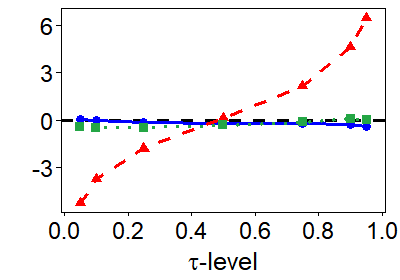}}
&
\adjustbox{valign=m}{%
  \includegraphics[width=0.31\textwidth]
  {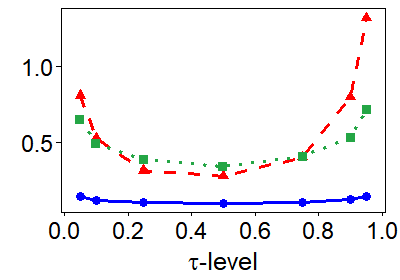}}
&
\adjustbox{valign=m}{%
  \includegraphics[width=0.31\textwidth]
  {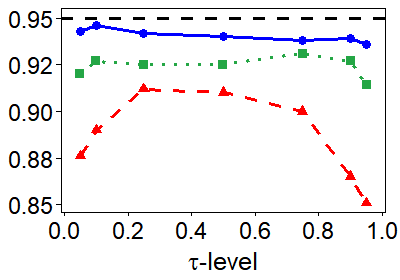}}
\\
\noalign{\vskip 2mm}

% DGP 3
\adjustbox{valign=m}{\rotatebox{90}{\textbf{DGP3}}}
&
\adjustbox{valign=m}{%
  \includegraphics[width=0.31\textwidth]
  {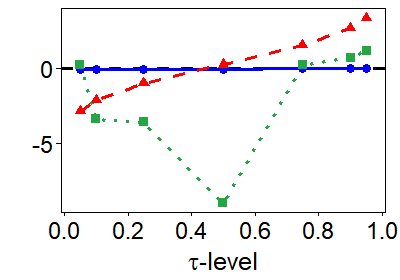}}
&
\adjustbox{valign=m}{%
  \includegraphics[width=0.31\textwidth]
  {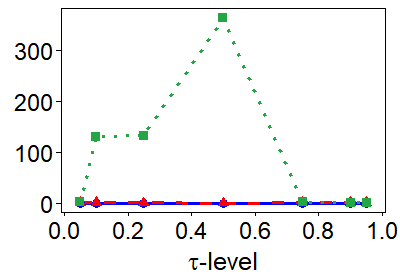}}
&
\adjustbox{valign=m}{%
  \includegraphics[width=0.31\textwidth]
  {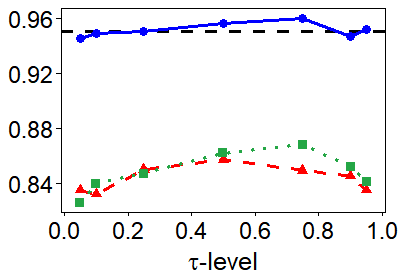}}
\\
\noalign{\vskip 2mm}

% DGP 4
\adjustbox{valign=m}{\rotatebox{90}{\textbf{DGP4}}}
&
\adjustbox{valign=m}{%
  \includegraphics[width=0.31\textwidth]
  {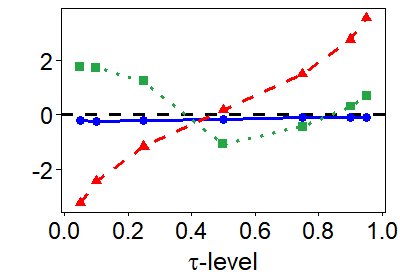}}
&
\adjustbox{valign=m}{%
  \includegraphics[width=0.31\textwidth]
  {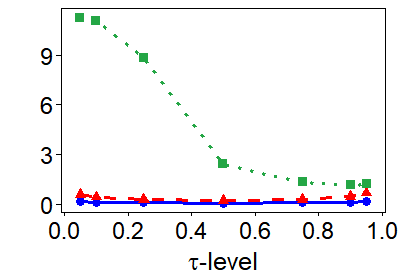}}
&
\adjustbox{valign=m}{%
  \includegraphics[width=0.31\textwidth]
  {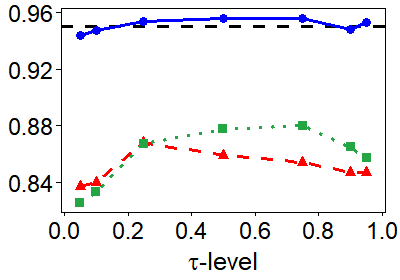}}
\\
\noalign{\vskip 3mm}

&
{\bfseries (a) Bias ($\times10^{-2}$)}
&
{\bfseries (b) MSE ($\times10^{-2}$)}
&
{\bfseries (c) CP}
\\
\noalign{\vskip 1mm}

\multicolumn{4}{c}{
  \includegraphics[
    width=0.80\textwidth
  ]{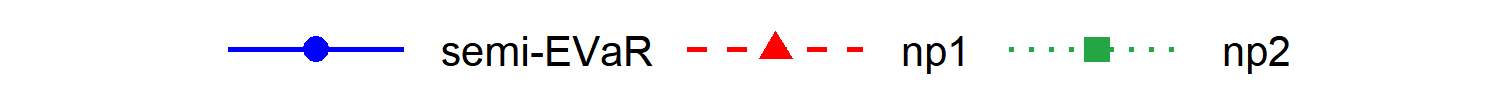}
}

\end{tabular}
\caption{
Simulation results for the proposed semiparametric estimator (“semi-EVaR”),
and two nonparametric estimators (``np1" and ``np2")
when $\varepsilon_t \sim N(0,1)$ and $n=500$. 
Bias: empirical bias; 
MSE: mean squared error; 
CP: 95\% coverage probability.
}
\label{fig:norm-n500}
\end{figure}

\begin{figure}%[htbp]
\centering
\setlength{\tabcolsep}{3pt}
\renewcommand{\arraystretch}{1}
\begin{tabular}{
@{}
c
ccc
@{}
}

% DGP 1
\adjustbox{valign=m}{\rotatebox{90}{\textbf{DGP1}}}
&
\adjustbox{valign=m}{%
  \includegraphics[width=0.31\textwidth]
  {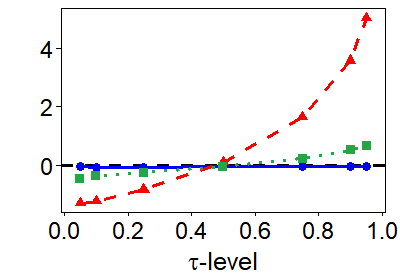}}
&
\adjustbox{valign=m}{%
  \includegraphics[width=0.31\textwidth]
  {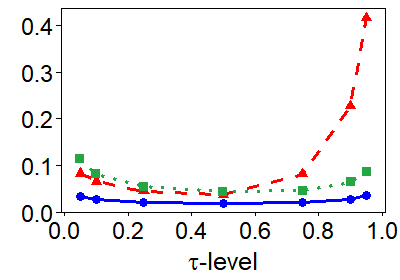}}
&
\adjustbox{valign=m}{%
  \includegraphics[width=0.31\textwidth]
  {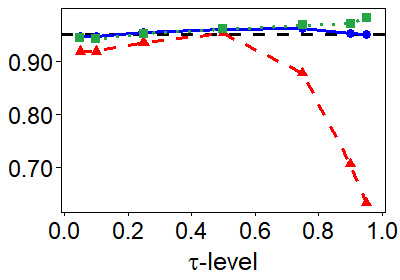}}
\\
\noalign{\vskip 2mm}

% DGP 2
\adjustbox{valign=m}{\rotatebox{90}{\textbf{DGP2}}}
&
\adjustbox{valign=m}{%
  \includegraphics[width=0.31\textwidth]
  {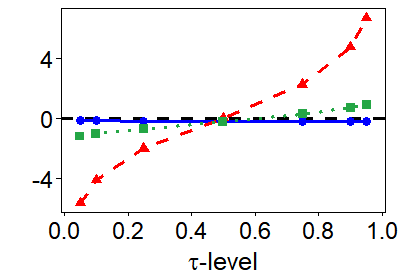}}
&
\adjustbox{valign=m}{%
  \includegraphics[width=0.31\textwidth]
  {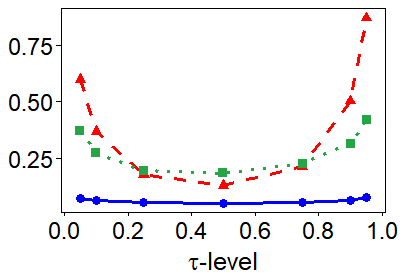}}
&
\adjustbox{valign=m}{%
  \includegraphics[width=0.31\textwidth]
  {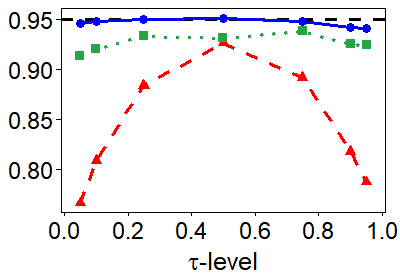}}
\\
\noalign{\vskip 2mm}

% DGP 3
\adjustbox{valign=m}{\rotatebox{90}{\textbf{DGP3}}}
&
\adjustbox{valign=m}{%
  \includegraphics[width=0.31\textwidth]
  {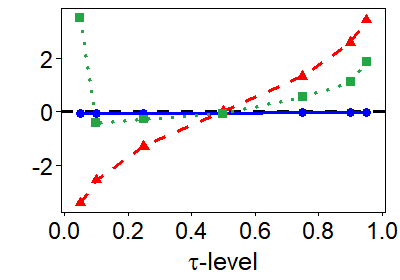}}
&
\adjustbox{valign=m}{%
  \includegraphics[width=0.31\textwidth]
  {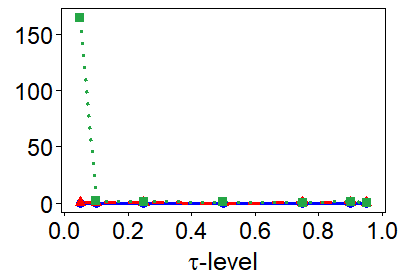}}
&
\adjustbox{valign=m}{%
  \includegraphics[width=0.31\textwidth]
  {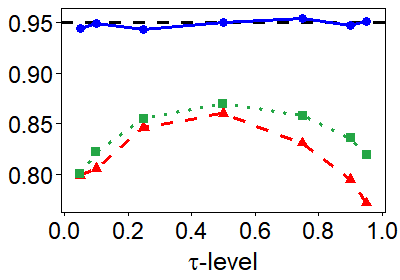}}
\\
\noalign{\vskip 2mm}

% DGP 4
\adjustbox{valign=m}{\rotatebox{90}{\textbf{DGP4}}}
&
\adjustbox{valign=m}{%
  \includegraphics[width=0.31\textwidth]
  {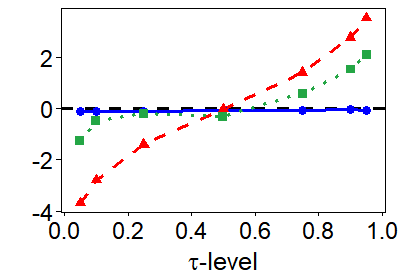}}
&
\adjustbox{valign=m}{%
  \includegraphics[width=0.31\textwidth]
  {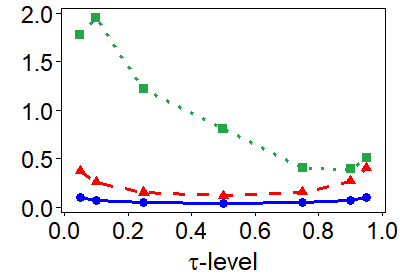}}
&
\adjustbox{valign=m}{%
  \includegraphics[width=0.31\textwidth]
  {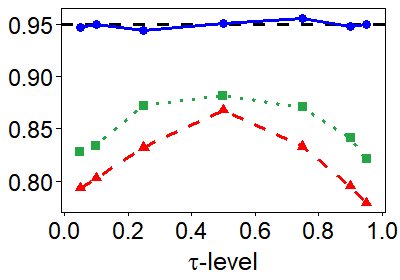}}
\\
\noalign{\vskip 3mm}

&
{\bfseries (a) Bias ($\times10^{-2}$)}
&
{\bfseries (b) MSE ($\times10^{-2}$)}
&
{\bfseries (c) CP}
\\
\noalign{\vskip 1mm}

\multicolumn{4}{c}{
  \includegraphics[
    width=0.80\textwidth
  ]{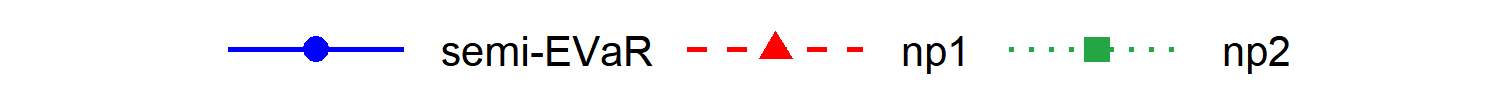}
}

\end{tabular}
\caption{
Simulation results for the proposed semiparametric estimator (“semi-EVaR”),
and two nonparametric estimators (``np1" and ``np2")
when $\varepsilon_t \sim N(0,1)$ and $n=1000$. 
Bias: empirical bias; 
MSE: mean squared error; 
CP: 95\% coverage probability.
}
\label{fig:norm-n1000}
\end{figure}

\begin{figure}%[htbp]
\centering
\setlength{\tabcolsep}{3pt}
\renewcommand{\arraystretch}{1}
\begin{tabular}{
@{}
c
ccc
@{}
}

% DGP 1
\adjustbox{valign=m}{\rotatebox{90}{\textbf{DGP1}}}
&
\adjustbox{valign=m}{%
  \includegraphics[width=0.31\textwidth]
  {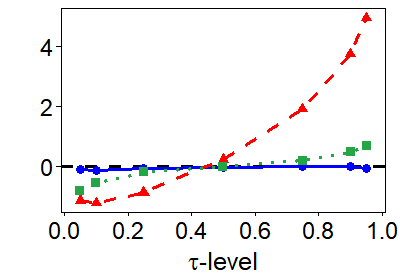}}
&
\adjustbox{valign=m}{%
  \includegraphics[width=0.31\textwidth]
  {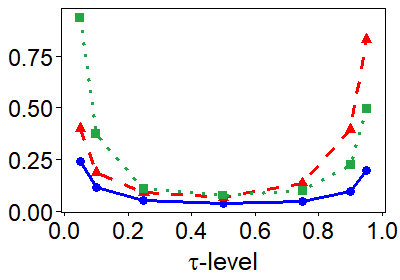}}
&
\adjustbox{valign=m}{%
  \includegraphics[width=0.31\textwidth]
  {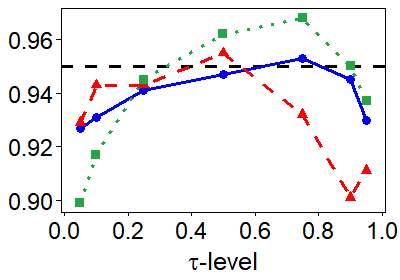}}
\\
\noalign{\vskip 2mm}

% DGP 2
\adjustbox{valign=m}{\rotatebox{90}{\textbf{DGP2}}}
&
\adjustbox{valign=m}{%
  \includegraphics[width=0.31\textwidth]
  {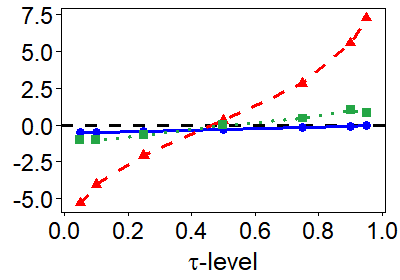}}
&
\adjustbox{valign=m}{%
  \includegraphics[width=0.31\textwidth]
  {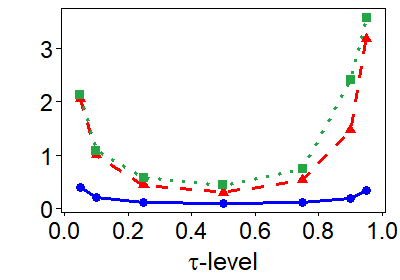}}
&
\adjustbox{valign=m}{%
  \includegraphics[width=0.31\textwidth]
  {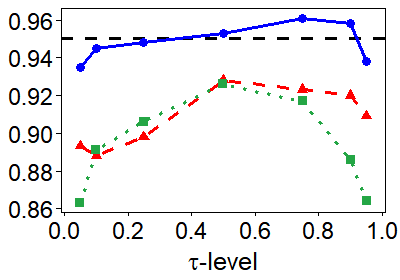}}
\\
\noalign{\vskip 2mm}

% DGP 3
\adjustbox{valign=m}{\rotatebox{90}{\textbf{DGP3}}}
&
\adjustbox{valign=m}{%
  \includegraphics[width=0.31\textwidth]
  {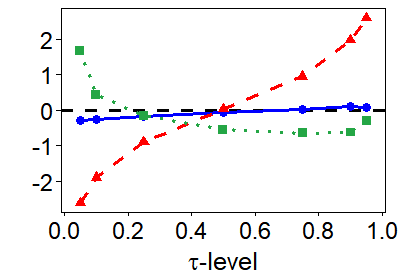}}
&
\adjustbox{valign=m}{%
  \includegraphics[width=0.31\textwidth]
  {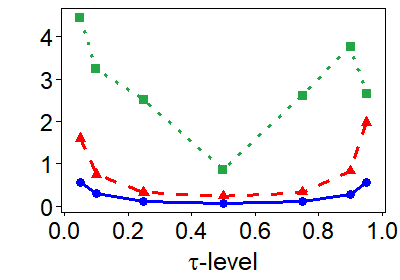}}
&
\adjustbox{valign=m}{%
  \includegraphics[width=0.31\textwidth]
  {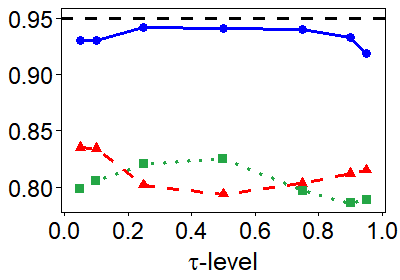}}
\\
\noalign{\vskip 2mm}

% DGP 4
\adjustbox{valign=m}{\rotatebox{90}{\textbf{DGP4}}}
&
\adjustbox{valign=m}{%
  \includegraphics[width=0.31\textwidth]
  {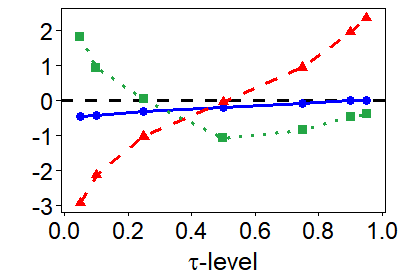}}
&
\adjustbox{valign=m}{%
  \includegraphics[width=0.31\textwidth]
  {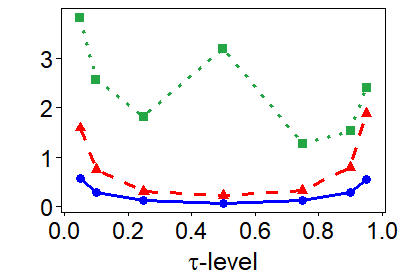}}
&
\adjustbox{valign=m}{%
  \includegraphics[width=0.31\textwidth]
  {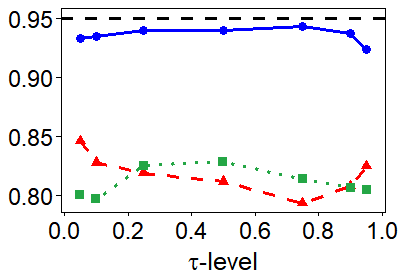}}
\\
\noalign{\vskip 3mm}

&
{\bfseries (a) Bias ($\times10^{-2}$)}
&
{\bfseries (b) MSE ($\times10^{-2}$)}
&
{\bfseries (c) CP}
\\
\noalign{\vskip 1mm}

\multicolumn{4}{c}{
  \includegraphics[
    width=0.80\textwidth
  ]{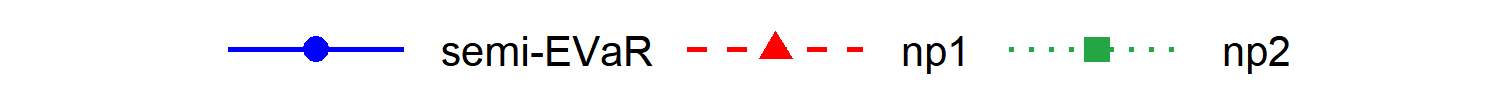}
}

\end{tabular}
\caption{
Simulation results for the proposed semiparametric estimator (“semi-EVaR”),
and two nonparametric estimators (``np1" and ``np2")
when $\varepsilon_t \sim t_4/\sqrt{2}$ and $n=500$. 
Bias: empirical bias; 
MSE: mean squared error; 
CP: 95\% coverage probability.
}
\label{fig:t-n500}
\end{figure}

\begin{figure}%[htbp]
\centering
\setlength{\tabcolsep}{3pt}
\renewcommand{\arraystretch}{1}
\begin{tabular}{
@{}
c
ccc
@{}
}

% DGP 1
\adjustbox{valign=m}{\rotatebox{90}{\textbf{DGP1}}}
&
\adjustbox{valign=m}{%
  \includegraphics[width=0.31\textwidth]
  {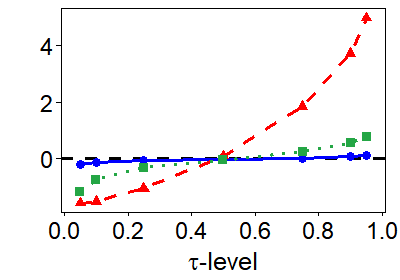}}
&
\adjustbox{valign=m}{%
  \includegraphics[width=0.31\textwidth]
  {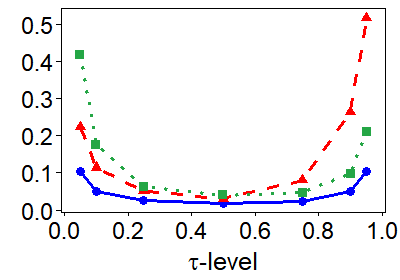}}
&
\adjustbox{valign=m}{%
  \includegraphics[width=0.31\textwidth]
  {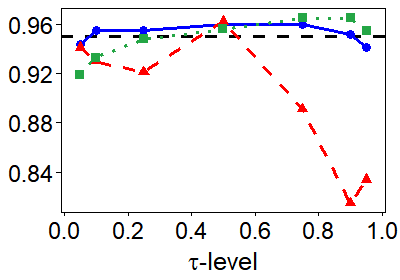}}
\\
\noalign{\vskip 2mm}

% DGP 2
\adjustbox{valign=m}{\rotatebox{90}{\textbf{DGP2}}}
&
\adjustbox{valign=m}{%
  \includegraphics[width=0.31\textwidth]
  {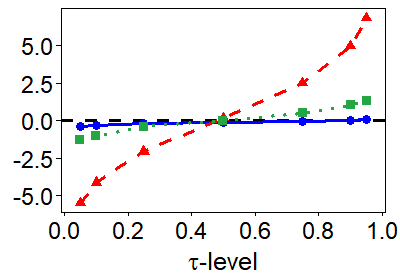}}
&
\adjustbox{valign=m}{%
  \includegraphics[width=0.31\textwidth]
  {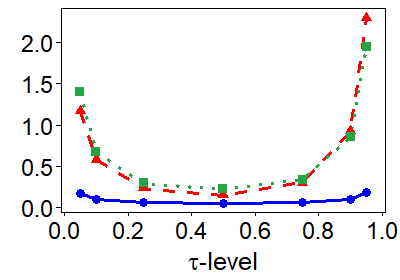}}
&
\adjustbox{valign=m}{%
  \includegraphics[width=0.31\textwidth]
  {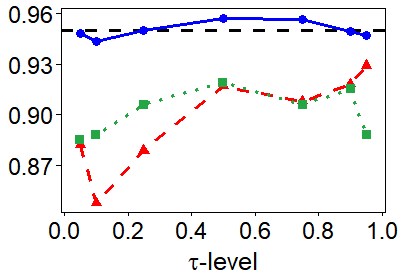}}
\\
\noalign{\vskip 2mm}

% DGP 3
\adjustbox{valign=m}{\rotatebox{90}{\textbf{DGP3}}}
&
\adjustbox{valign=m}{%
  \includegraphics[width=0.31\textwidth]
  {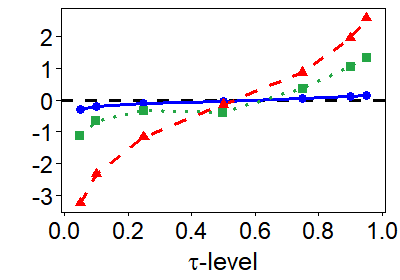}}
&
\adjustbox{valign=m}{%
  \includegraphics[width=0.31\textwidth]
  {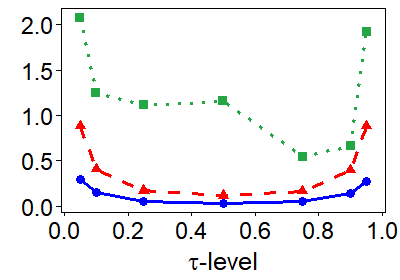}}
&
\adjustbox{valign=m}{%
  \includegraphics[width=0.31\textwidth]
  {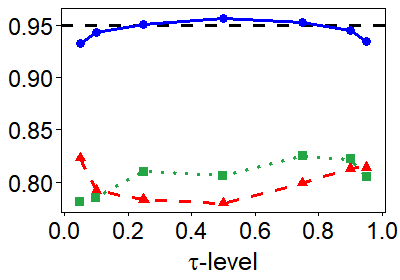}}
\\
\noalign{\vskip 2mm}

% DGP 4
\adjustbox{valign=m}{\rotatebox{90}{\textbf{DGP4}}}
&
\adjustbox{valign=m}{%
  \includegraphics[width=0.31\textwidth]
  {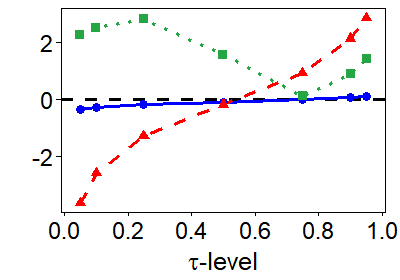}}
&
\adjustbox{valign=m}{%
  \includegraphics[width=0.31\textwidth]
  {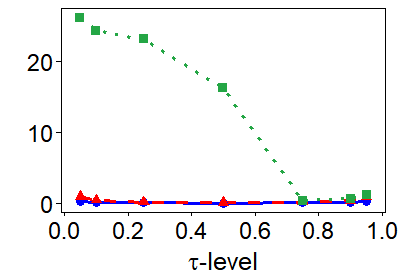}}
&
\adjustbox{valign=m}{%
  \includegraphics[width=0.31\textwidth]
  {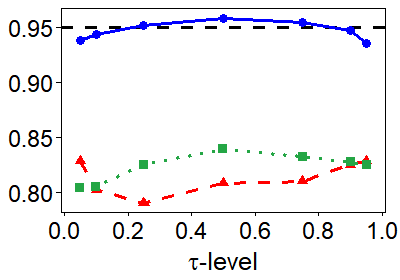}}
\\
\noalign{\vskip 3mm}

&
{\bfseries (a) Bias ($\times10^{-2}$)}
&
{\bfseries (b) MSE ($\times10^{-2}$)}
&
{\bfseries (c) CP}
\\
\noalign{\vskip 1mm}

\multicolumn{4}{c}{
  \includegraphics[
    width=0.80\textwidth
  ]{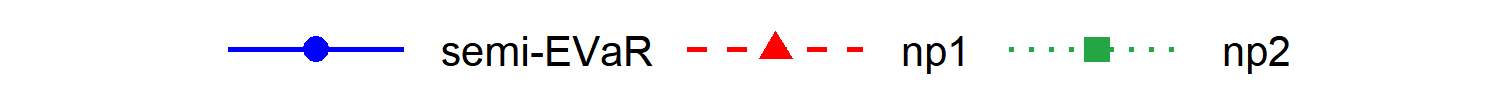}
}

\end{tabular}
\caption{
Simulation results for the proposed semiparametric estimator (“semi-EVaR”),
and two nonparametric estimators (``np1" and ``np2")
when $\varepsilon_t \sim t_4/\sqrt{2}$ and $n=1000$. 
Bias: empirical bias; 
MSE: mean squared error; 
CP: 95\% coverage probability.
}
\label{fig:t-n1000}
\end{figure}

The true EVaR values are obtained from the known stationary distribution of the errors.
Confidence intervals are constructed using the bootstrap procedure described in Section~\ref{ss:implement}, with $M=1000$ bootstrap replications.
Based on 1000 Monte Carlo replications, 
Figures~\ref{fig:norm-n500}-- \ref{fig:t-n1000} present the empirical bias (``Bias''), mean squared error (``MSE''), and 95\% coverage probability (``CP'') for four DGPs under normal and Student-$t$ error distributions, respectively. 
Detailed numerical results are provided in the Appendix.
For comparison, we also report results from two nonparametric estimators: (i) the estimator obtained by inverting the local polynomial estimator of the conditional loss-gain function (``np1'')  in \citet{zhang2025nonparametric}, and (ii) the local linear expectile regression estimator (``np2'') proposed by \citet{yao1996asymmetric}.

The results illustrate that the proposed semiparametric estimator (``semi-EVaR") performs consistently well across all scenarios considered. 
First, the empirical biases are uniformly small, suggesting that the proposed estimator is asymptotically unbiased. 
Second, the finite-sample coverage probabilities are close to the nominal 95\% level, providing empirical support for the validity of the asymptotic distribution.
Third, the proposed estimator generally achieves smaller absolute Biases and lower MSEs, as well as more accurate coverage probabilities, compared to the two nonparametric estimators (``np1" and ``np2") in nearly all settings.
Notably, 
under the Student-$t$ error distribution, 
which has heavier tails than the normal distribution, the proposed estimator remains robust, whereas the alternative estimators sometimes exhibit undercoverage, 
particularly for the GARCH-type models (DGPs 3 and 4).
Furthermore, increasing the sample size $n$ from $500$ to $1000$ leads to a reduction in MSE across all methods. 
As the $\tau$-level approaches $0.5$, the estimates tend to exhibit lower variability and smaller MSEs, reflecting the reduced difficulty of estimation at less extreme expectile levels.
To sum up, the proposed semiparametric estimator is superior to the two nonparametric estimators.

\subsection{Comparison with completely parametric estimator}

Our proposed semiparametric estimator uses the sample expectile of the fitted residuals in \eqref{eqn:EVaR-e}, without imposing a parametric distribution on the errors. 
A natural question is how much efficiency could be lost by leaving the error distribution unspecified, 
relative to a completely parametric estimator that exploits knowledge of the error distribution. 
%To investigate this issue, 
For comparison,  
we consider the completely parametric estimator
\[
\widetilde{\operatorname{EVaR}}_\tau(\bx)
=
m(\bx,\widetilde{\bbeta})
+
\sigma(\bx,\widetilde{\bbeta})\mu_\tau(e_t),
\]
where 
$\widetilde{\bbeta}$ is the maximum likelihood estimator of $\bbeta$
when the distribution of $e_t$ is specified, 
and 
$\mu_\tau(e_t)$ is directly calculated from the specified  distribution of $e_t$. 
We consider two parametric distributions for the error $e_t$:
(i) the standard normal distribution $N(0,1)$ and
(ii) the standardized Student-$t$ distribution $t_4/\sqrt{2}$. 
Their parametric estimations are denoted by ``p-norm" and ``p-t", respectively.

To compare their performance, 
we define the relative mean squared error (RMSE) of the completely parametric estimator $\widetilde{\operatorname{EVaR}}_\tau(\bx)$ relative to the semiparametric estimator $\widehat{\operatorname{EVaR}}_\tau(\bx)$ as
$
\mathrm{RMSE}
=
\frac{
\mathrm{MSE}\left(
\widetilde{\operatorname{EVaR}}_\tau(\bx)
\right)
}{
\mathrm{MSE}\left(
\widehat{\operatorname{EVaR}}_\tau(\bx)
\right)
}.
$
Thus, $\mathrm{RMSE}$ greater than one indicates an efficiency advantage of the proposed semiparametric estimator with better MSE performance.

Table~\ref{tab:parametric-comparison} reports the RMSE results under both correctly specified and misspecified error distributions.
As expected, 
when the error distribution is correctly specified, the completely parametric estimator generally has a smaller MSE. 
However, once the error distribution is misspecified,  
the completely parametric estimator may perform substantially poor, 
as the RMSE is greater than one in most cases. 
Particularly, 
when a Student-$t$ distribution is incorrectly assumed for Gaussian errors at $\tau=0.05$ and $0.95$ with $n=1000$, 
the MSE of the misspecified parametric estimator can be more than 10 times that of the proposed semiparametric estimator. 
Since the error distribution is usually unknown in practice, 
we thus recommend the robust proposed semiparametric estimator 
against potential misspecification of the error distribution, 
retaining only a slight efficiency loss relative to the correctly specific parametric estimator. 
%which  incurs only a slight efficiency loss relative to the completely parametric estimator under correct distributional specification, while providing greater robustness against potential misspecification of the error distribution.
%Therefore, the efficiency advantage of the parametric approach relies on the strong assumption of correct distributional specification, whereas in practice the error distribution is usually unknown.

\begin{table}
\centering
\footnotesize
\renewcommand{\arraystretch}{1.2}
\setlength{\tabcolsep}{0.5mm}

\caption{Relative MSE (RMSE) of the completely parametric estimator compared with the proposed semiparametric estimator.
$\mathrm{RMSE} > 1$ indicates better performance of the proposed semiparametric estimator.
}
\label{tab:parametric-comparison}

\makebox[\textwidth][c]{%
\begin{tabular}{@{}ccc @{\hspace{3mm}} c *{7}{r} @{\hspace{4mm}} *{7}{r}@{}}

\hline
& & & &
\multicolumn{7}{c}{$n=500$} &
\multicolumn{7}{c}{$n=1000$} \\
\cmidrule(lr){5-11}
\cmidrule(lr){12-18}

DGPs & Error & Method & $\tau$
& 0.05 & 0.10 & 0.25 & 0.50 & 0.75 & 0.90 & 0.95
& 0.05 & 0.10 & 0.25 & 0.50 & 0.75 & 0.90 & 0.95 \\
\hline

DGP 1 & $N(0,1)$ & p-norm &
& 0.866 & 0.937 & 0.970 & 0.968 & 0.964 & 0.924 & 0.849
& 0.808 & 0.899 & 0.959 & 0.969 & 0.977 & 0.942 & 0.865 \\

& & p-t &
& 8.803 & 3.502 & 1.327 & 1.140 & 1.417 & 3.972 & 9.988
& 18.628 & 6.891 & 1.693 & 1.135 & 1.599 & 6.503 & 18.029 \\

& $t_4/\sqrt{2}$ & p-norm &
& 0.875 & 1.425 & 1.755 & 0.989 & 1.672 & 1.466 & 0.946
& 0.831 & 1.680 & 2.269 & 0.965 & 2.237 & 1.660 & 0.811 \\

& & p-t &
& 0.268 & 0.414 & 0.629 & 0.724 & 0.663 & 0.463 & 0.310
& 0.312 & 0.457 & 0.654 & 0.743 & 0.680 & 0.481 & 0.323 \\

\addlinespace

DGP 2 & $N(0,1)$ & p-norm &
& 0.893 & 0.950 & 0.975 & 0.975 & 0.970 & 0.943 & 0.896
& 0.863 & 0.924 & 0.961 & 0.969 & 0.970 & 0.943 & 0.885 \\

& & p-t &
& 6.679 & 2.783 & 1.270 & 1.100 & 1.172 & 2.465 & 6.158
& 12.869 & 4.691 & 1.450 & 1.086 & 1.243 & 3.889 & 11.217 \\

& $t_4/\sqrt{2}$ & p-norm &
& 0.833 & 1.304 & 1.497 & 0.985 & 1.294 & 1.170 & 0.804
& 0.881 & 1.542 & 1.757 & 0.972 & 1.625 & 1.434 & 0.850 \\

& & p-t &
& 0.311 & 0.473 & 0.662 & 0.720 & 0.682 & 0.516 & 0.360
& 0.393 & 0.546 & 0.691 & 0.727 & 0.684 & 0.521 & 0.360 \\

\addlinespace

DGP 3 & $N(0,1)$ & p-norm &
& 0.853 & 0.878 & 0.869 & 0.854 & 0.866 & 0.866 & 0.832
& 0.850 & 0.872 & 0.868 & 0.846 & 0.869 & 0.881 & 0.854 \\

& & p-t &
& 6.687 & 3.040 & 1.236 & 0.973 & 1.199 & 2.952 & 6.488
& 11.817 & 4.784 & 1.423 & 0.972 & 1.383 & 4.638 & 11.430 \\

& $t_4/\sqrt{2}$ & p-norm &
& 0.852 & 1.254 & 1.745 & 1.375 & 1.659 & 1.233 & 0.862
& 0.926 & 1.540 & 2.305 & 1.496 & 2.124 & 1.467 & 0.910 \\

& & p-t &
& 0.342 & 0.433 & 0.587 & 0.688 & 0.565 & 0.422 & 0.338
& 0.322 & 0.410 & 0.565 & 0.690 & 0.577 & 0.428 & 0.339 \\

\addlinespace

DGP 4 & $N(0,1)$ & p-norm &
& 0.869 & 0.894 & 0.892 & 0.880 & 0.888 & 0.881 & 0.846
& 0.872 & 0.893 & 0.890 & 0.868 & 0.884 & 0.890 & 0.861 \\

& & p-t &
& 6.592 & 3.074 & 1.309 & 1.014 & 1.160 & 2.728 & 6.074
& 11.477 & 4.739 & 1.502 & 1.017 & 1.323 & 4.289 & 10.701 \\

& $t_4/\sqrt{2}$ & p-norm &
& 0.818 & 1.267 & 1.800 & 1.207 & 1.360 & 1.121 & 0.853
& 0.887 & 1.569 & 2.387 & 1.287 & 1.720 & 1.309 & 0.893 \\

& & p-t &
& 0.345 & 0.436 & 0.593 & 0.697 & 0.581 & 0.438 & 0.349
& 0.333 & 0.425 & 0.583 & 0.702 & 0.590 & 0.442 & 0.351 \\

\hline
\end{tabular}
}
\end{table}

\subsection{Performance under model misspecification}
\label{ss:misspec}

In this subsection, we evaluate the performance of the proposed semiparametric estimator under model misspecification. 
The data are generated from the GJR-GARCH model in DGP 4 with 
$(\mu,\omega,\alpha,\beta)=(0.5,0.2,0.4,0.2)$ and $\gamma \in \{0.2,0.5\}$, while the misspecified model used for estimation is the GARCH model in DGP 3. 
The parameter $\gamma$ controls the degree of deviation between the true and misspecified models. 
The results for the proposed semiparametric estimator are reported in Table~\ref{tab:misspec}, with the sample size set to $n=500$. 
For comparison, we also report the results for the two nonparametric estimators and the two completely parametric estimators.
The results show that the proposed semiparametric estimator remains robust under model misspecification and delivers better overall performance than the competing methods.

%\newgeometry{left=1mm,right=1mm}
\begin{center}
\footnotesize
\renewcommand{\arraystretch}{0.7}
\setlength{\tabcolsep}{0.2mm}
\setlength{\LTleft}{\dimexpr-1.8cm\relax}
\setlength{\LTright}{\dimexpr-1.8cm\relax}

\begin{longtable}{@{}ccc c *{7}{r} @{\hspace{4mm}} *{7}{r}@{}}

\caption{Simulation results in the presence of model misspecification with $n=500$.
Bias ($\times10^{-2}$): the empirical bias; SD: the empirical standard deviation; 
ESE: the average estimated standard error; CP: the 95\% coverage probability; 
MSE ($\times10^{-2}$): the mean squared error.}
\label{tab:misspec}\\

\hline
& & & &
\multicolumn{7}{c}{$\gamma=0.2$} &
\multicolumn{7}{c}{$\gamma=0.5$}\\
\cmidrule(lr){5-11}
\cmidrule(lr){12-18}

Error & Method & & $\tau$
& 0.05 & 0.10 & 0.25 & 0.50 & 0.75 & 0.90 & 0.95
& 0.05 & 0.10 & 0.25 & 0.50 & 0.75 & 0.90 & 0.95 \\
\hline

\endfirsthead

\multicolumn{18}{c}{{\bfseries \tablename\ \thetable{} -- Continued.}}\\
\hline

Error & Method & & $\tau$
& 0.05 & 0.10 & 0.25 & 0.50 & 0.75 & 0.90 & 0.95
& 0.05 & 0.10 & 0.25 & 0.50 & 0.75 & 0.90 & 0.95 \\
\hline

\endhead

\multicolumn{18}{r}{(Continued)}\\
\hline
\endfoot

\hline
\endlastfoot

%====================================================
% Normal errors
%====================================================

$N(0,1)$ & & True &
& -0.124 & 0.028 & 0.261 & 0.500 & 0.739 & 0.972 & 1.124
& -0.124 & 0.028 & 0.261 & 0.500 & 0.739 & 0.972 & 1.124 \\

& semi-EVaR & Bias &
& 0.226 & 0.194 & 0.171 & 0.163 & 0.160 & 0.158 & 0.139
& 0.877 & 0.898 & 0.869 & 0.798 & 0.735 & 0.736 & 0.777 \\

& & SD &
& 0.045 & 0.038 & 0.031 & 0.028 & 0.031 & 0.038 & 0.045
& 0.047 & 0.040 & 0.032 & 0.030 & 0.033 & 0.041 & 0.048 \\

& & ESE &
& 0.045 & 0.039 & 0.032 & 0.029 & 0.032 & 0.038 & 0.045
& 0.049 & 0.042 & 0.034 & 0.031 & 0.033 & 0.041 & 0.048 \\

& & CP &
& 0.940 & 0.942 & 0.952 & 0.959 & 0.962 & 0.951 & 0.951
& 0.936 & 0.938 & 0.944 & 0.953 & 0.955 & 0.947 & 0.952 \\

& & MSE &
& 0.200 & 0.144 & 0.096 & 0.080 & 0.096 & 0.146 & 0.206
& 0.228 & 0.166 & 0.113 & 0.095 & 0.114 & 0.172 & 0.239 \\

\addlinespace[4pt]

& np1 & Bias &
& -1.550 & -1.162 & -0.472 & 0.224 & 0.931 & 1.577 & 1.910
& -2.408 & -1.850 & -0.850 & 0.227 & 1.257 & 2.351 & 2.990 \\

& & SD &
& 0.069 & 0.060 & 0.051 & 0.048 & 0.052 & 0.064 & 0.084
& 0.070 & 0.060 & 0.051 & 0.047 & 0.051 & 0.062 & 0.076 \\

& & ESE &
& 0.054 & 0.046 & 0.039 & 0.037 & 0.040 & 0.048 & 0.060
& 0.059 & 0.050 & 0.043 & 0.041 & 0.043 & 0.052 & 0.062 \\

& & CP &
& 0.859 & 0.855 & 0.866 & 0.863 & 0.874 & 0.853 & 0.855
& 0.880 & 0.889 & 0.900 & 0.904 & 0.887 & 0.877 & 0.875 \\

& & MSE &
& 0.502 & 0.370 & 0.258 & 0.230 & 0.274 & 0.432 & 0.739
& 0.546 & 0.396 & 0.264 & 0.225 & 0.279 & 0.439 & 0.664 \\

\addlinespace[4pt]

& np2 & Bias &
& 0.547 & 0.311 & 0.040 & -0.075 & -0.293 & 0.179 & 0.496
& 1.533 & 2.286 & 0.921 & -0.303 & -0.026 & 0.551 & 0.924 \\

& & SD &
& 0.180 & 0.170 & 0.162 & 0.149 & 0.075 & 0.082 & 0.084
& 0.177 & 0.204 & 0.251 & 0.230 & 0.090 & 0.085 & 0.094 \\

& & ESE &
& 0.074 & 0.067 & 0.062 & 0.055 & 0.049 & 0.053 & 0.061
& 0.087 & 0.086 & 0.087 & 0.075 & 0.059 & 0.063 & 0.069 \\

& & CP &
& 0.844 & 0.835 & 0.864 & 0.856 & 0.865 & 0.858 & 0.867
& 0.854 & 0.854 & 0.883 & 0.885 & 0.898 & 0.893 & 0.884 \\

& & MSE &
& 3.231 & 2.900 & 2.616 & 2.210 & 0.564 & 0.664 & 0.712
& 3.170 & 4.220 & 6.299 & 5.287 & 0.811 & 0.726 & 0.889 \\

\addlinespace[4pt]

& p-norm & Bias &
& 1.224 & 1.219 & 1.211 & 1.203 & 1.195 & 1.187 & 1.182
& 3.219 & 3.215 & 3.207 & 3.200 & 3.192 & 3.185 & 3.180 \\

& & SD &
& 0.042 & 0.036 & 0.029 & 0.027 & 0.029 & 0.036 & 0.041
& 0.043 & 0.037 & 0.030 & 0.028 & 0.031 & 0.038 & 0.044 \\

& & ESE &
& 0.041 & 0.036 & 0.029 & 0.027 & 0.030 & 0.036 & 0.042
& 0.043 & 0.037 & 0.030 & 0.028 & 0.031 & 0.038 & 0.044 \\

& & CP &
& 0.929 & 0.922 & 0.920 & 0.935 & 0.941 & 0.946 & 0.952
& 0.851 & 0.826 & 0.794 & 0.781 & 0.846 & 0.885 & 0.905 \\

& & MSE &
& 0.189 & 0.145 & 0.101 & 0.086 & 0.099 & 0.141 & 0.184
& 0.289 & 0.242 & 0.196 & 0.180 & 0.198 & 0.247 & 0.296 \\

\addlinespace[4pt]

& p-t & Bias &
& -10.174 & -4.721 & -0.719 & 0.751 & 2.221 & 6.223 & 11.676
& -8.914 & -3.430 & 0.612 & 2.119 & 3.627 & 7.669 & 13.152 \\

& & SD &
& 0.050 & 0.041 & 0.032 & 0.028 & 0.031 & 0.039 & 0.047
& 0.052 & 0.043 & 0.034 & 0.030 & 0.033 & 0.042 & 0.050 \\

& & ESE &
& 0.056 & 0.045 & 0.032 & 0.027 & 0.032 & 0.045 & 0.056
& 0.059 & 0.047 & 0.034 & 0.029 & 0.034 & 0.047 & 0.059 \\

& & CP &
& 0.579 & 0.852 & 0.947 & 0.942 & 0.922 & 0.767 & 0.459
& 0.708 & 0.931 & 0.939 & 0.871 & 0.854 & 0.679 & 0.384 \\

& & MSE &
& 1.280 & 0.390 & 0.108 & 0.086 & 0.145 & 0.540 & 1.587
& 1.067 & 0.303 & 0.117 & 0.134 & 0.238 & 0.761 & 1.984 \\

\addlinespace[6pt]

%====================================================
% Student-t errors
%====================================================

$t_4/\sqrt{2}$ & & True &
& -0.123 & 0.053 & 0.285 & 0.500 & 0.715 & 0.947 & 1.123
& -0.123 & 0.053 & 0.285 & 0.500 & 0.715 & 0.947 & 1.123 \\

& semi-EVaR & Bias &
& 0.000 & -0.004 & 0.020 & 0.041 & 0.068 & 0.083 & 0.038
& 0.287 & 0.305 & 0.345 & 0.343 & 0.350 & 0.390 & 0.389 \\

& & SD &
& 0.078 & 0.055 & 0.035 & 0.027 & 0.035 & 0.055 & 0.077
& 0.080 & 0.057 & 0.036 & 0.029 & 0.036 & 0.056 & 0.079 \\

& & ESE &
& 0.071 & 0.051 & 0.034 & 0.027 & 0.033 & 0.051 & 0.071
& 0.073 & 0.053 & 0.035 & 0.029 & 0.034 & 0.052 & 0.072 \\

& & CP &
& 0.928 & 0.938 & 0.939 & 0.944 & 0.949 & 0.939 & 0.923
& 0.921 & 0.930 & 0.938 & 0.942 & 0.945 & 0.938 & 0.923 \\

& & MSE &
& 0.613 & 0.307 & 0.121 & 0.075 & 0.119 & 0.300 & 0.598
& 0.645 & 0.328 & 0.133 & 0.083 & 0.128 & 0.313 & 0.619 \\

\addlinespace[4pt]

& np1 & Bias &
& -1.295 & -0.998 & -0.474 & 0.007 & 0.531 & 1.107 & 1.520
& -2.151 & -1.557 & -0.702 & 0.086 & 0.857 & 1.804 & 2.367 \\

& & SD &
& 0.126 & 0.086 & 0.058 & 0.051 & 0.060 & 0.092 & 0.142
& 0.122 & 0.083 & 0.054 & 0.047 & 0.054 & 0.084 & 0.128 \\

& & ESE &
& 0.079 & 0.055 & 0.037 & 0.032 & 0.037 & 0.058 & 0.089
& 0.084 & 0.058 & 0.039 & 0.033 & 0.039 & 0.060 & 0.090 \\

& & CP &
& 0.825 & 0.820 & 0.797 & 0.775 & 0.782 & 0.800 & 0.800
& 0.847 & 0.856 & 0.849 & 0.844 & 0.853 & 0.862 & 0.862 \\

& & MSE &
& 1.594 & 0.757 & 0.336 & 0.258 & 0.358 & 0.863 & 2.029
& 1.541 & 0.720 & 0.301 & 0.218 & 0.299 & 0.738 & 1.702 \\

\addlinespace[4pt]

& np2 & Bias &
& 1.343 & 0.598 & 0.070 & -0.388 & -0.211 & -0.064 & -0.150
& 2.260 & 1.120 & 0.436 & -0.413 & 1.280 & 1.573 & 0.144 \\

& & SD &
& 0.200 & 0.156 & 0.112 & 0.097 & 0.084 & 0.111 & 0.148
& 0.283 & 0.175 & 0.150 & 0.114 & 0.429 & 0.425 & 0.137 \\

& & ESE &
& 0.109 & 0.087 & 0.065 & 0.052 & 0.051 & 0.066 & 0.090
& 0.125 & 0.097 & 0.074 & 0.059 & 0.067 & 0.081 & 0.094 \\

& & CP &
& 0.779 & 0.796 & 0.815 & 0.813 & 0.798 & 0.788 & 0.786
& 0.787 & 0.819 & 0.826 & 0.848 & 0.844 & 0.849 & 0.838 \\

& & MSE &
& 4.005 & 2.424 & 1.243 & 0.938 & 0.712 & 1.234 & 2.181
& 8.067 & 3.082 & 2.248 & 1.301 & 18.366 & 18.036 & 1.879 \\

\addlinespace[4pt]

& p-norm & Bias &
& 1.203 & -1.280 & -1.399 & 0.716 & 2.832 & 2.712 & 0.229
& 2.515 & 0.036 & -0.078 & 2.043 & 4.164 & 4.050 & 1.570 \\

& & SD &
& 0.072 & 0.058 & 0.040 & 0.031 & 0.039 & 0.057 & 0.071
& 0.074 & 0.060 & 0.042 & 0.034 & 0.042 & 0.059 & 0.073 \\

& & ESE &
& 0.038 & 0.033 & 0.028 & 0.025 & 0.028 & 0.034 & 0.039
& 0.039 & 0.034 & 0.028 & 0.026 & 0.028 & 0.034 & 0.039 \\

& & CP &
& 0.703 & 0.771 & 0.817 & 0.883 & 0.747 & 0.750 & 0.730
& 0.656 & 0.753 & 0.819 & 0.800 & 0.663 & 0.687 & 0.750 \\

& & MSE &
& 0.528 & 0.352 & 0.179 & 0.103 & 0.236 & 0.401 & 0.503
& 0.608 & 0.360 & 0.177 & 0.154 & 0.346 & 0.516 & 0.558 \\

\addlinespace[4pt]

& p-t & Bias &
& 0.212 & 0.252 & 0.305 & 0.354 & 0.403 & 0.455 & 0.495
& 0.780 & 0.841 & 0.921 & 0.996 & 1.071 & 1.151 & 1.212 \\

& & SD &
& 0.047 & 0.038 & 0.027 & 0.023 & 0.026 & 0.036 & 0.045
& 0.048 & 0.038 & 0.028 & 0.024 & 0.027 & 0.037 & 0.046 \\

& & ESE &
& 0.045 & 0.036 & 0.026 & 0.022 & 0.026 & 0.036 & 0.045
& 0.045 & 0.036 & 0.027 & 0.023 & 0.027 & 0.037 & 0.046 \\

& & CP &
& 0.933 & 0.930 & 0.936 & 0.934 & 0.940 & 0.949 & 0.952
& 0.920 & 0.914 & 0.913 & 0.921 & 0.937 & 0.950 & 0.951 \\

& & MSE &
& 0.221 & 0.141 & 0.076 & 0.054 & 0.071 & 0.132 & 0.208
& 0.234 & 0.154 & 0.088 & 0.067 & 0.085 & 0.147 & 0.224 \\

\bottomrule
\end{longtable}
\end{center}
%\restoregeometry

\subsection{Asymptotic versus bootstrap confidence intervals}
\label{ss:cp}
\begin{figure}%[htbp]
  \centering
  
  \begin{subfigure}{0.49\textwidth}
    \centering
    \includegraphics[width=\linewidth]{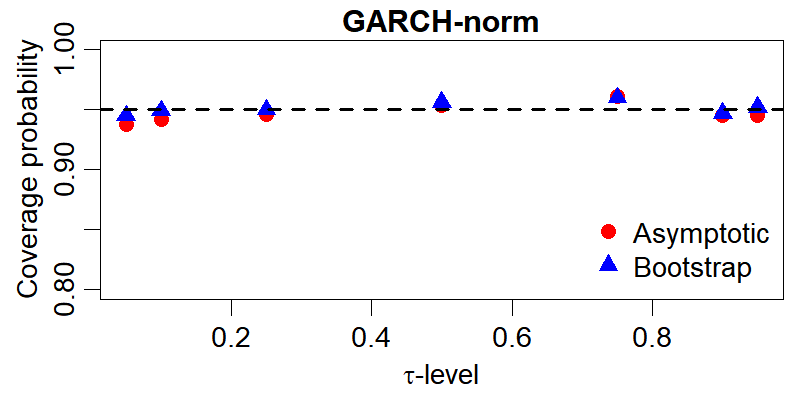}
  \end{subfigure}\hfill
  \begin{subfigure}{0.49\textwidth}
    \centering
    \includegraphics[width=\linewidth]{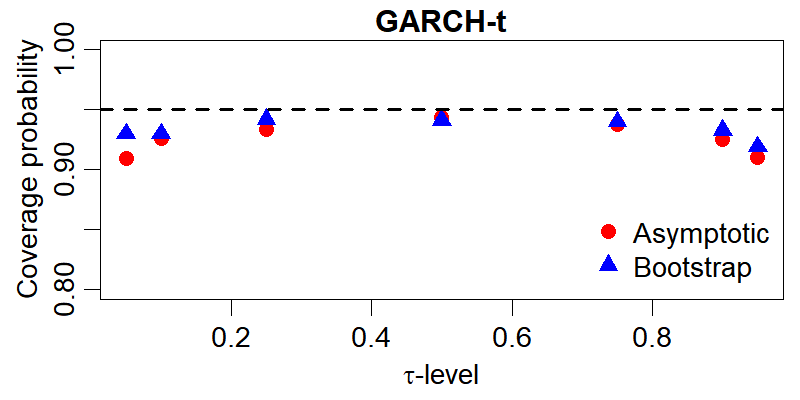}
  \end{subfigure}

  \vspace{0.5em}

  \begin{subfigure}{0.49\textwidth}
    \centering
    \includegraphics[width=\linewidth]{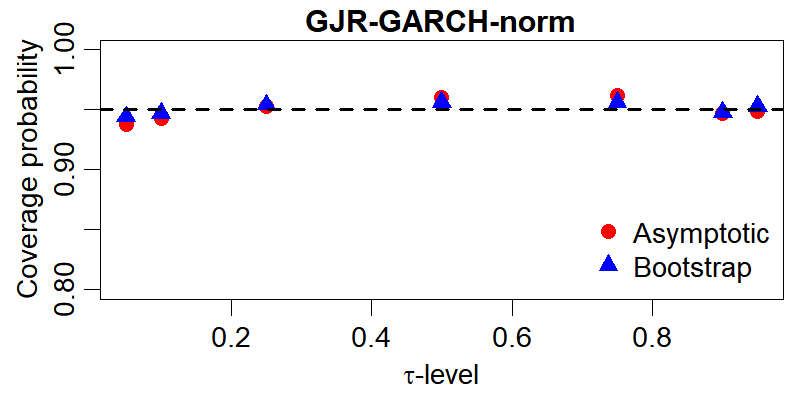}
  \end{subfigure}\hfill
  \begin{subfigure}{0.49\textwidth}
    \centering
    \includegraphics[width=\linewidth]{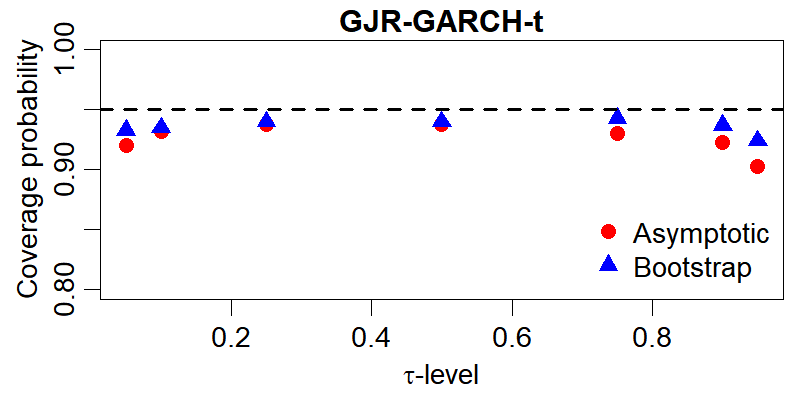}
  \end{subfigure}

  \caption{95\% empirical coverage probability of confidence interval when $n=500$.}
  \label{fig:cp_n500}
\end{figure}

\begin{figure}%[htbp]
  \centering
  
  \begin{subfigure}{0.49\textwidth}
    \centering
    \includegraphics[width=\linewidth]{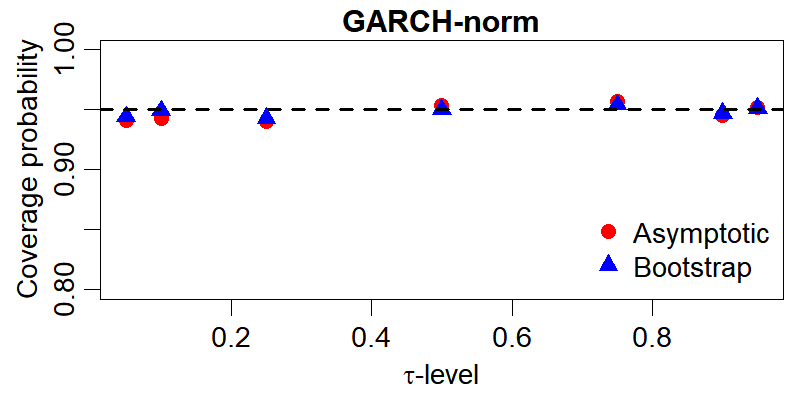}
  \end{subfigure}\hfill
  \begin{subfigure}{0.49\textwidth}
    \centering
    \includegraphics[width=\linewidth]{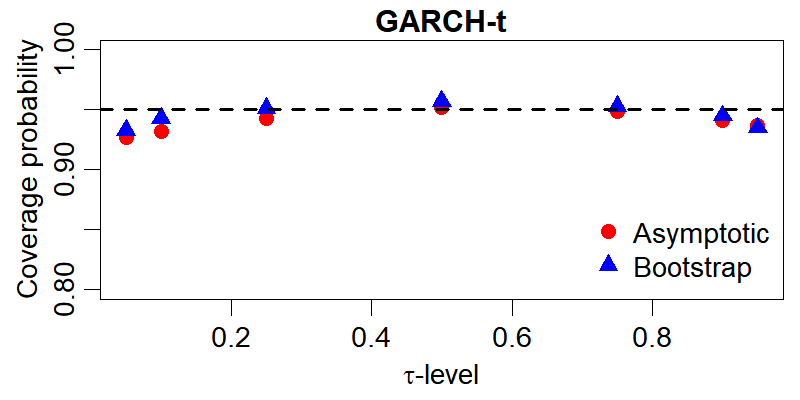}
  \end{subfigure}

  \vspace{0.5em}

  \begin{subfigure}{0.49\textwidth}
    \centering
    \includegraphics[width=\linewidth]{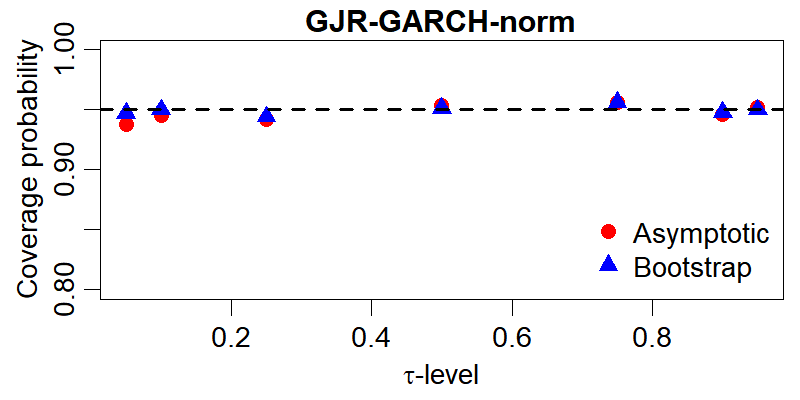}
  \end{subfigure}\hfill
  \begin{subfigure}{0.49\textwidth}
    \centering
    \includegraphics[width=\linewidth]{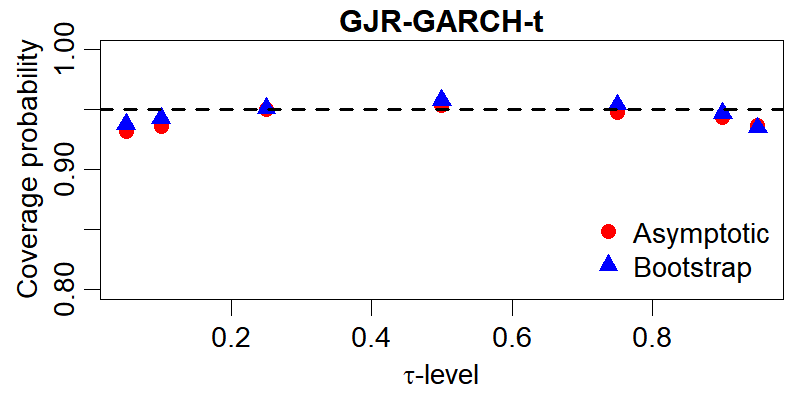}
  \end{subfigure}

  \caption{95\% empirical coverage probability of confidence interval when $n=1000$.}
  \label{fig:cp_n1000}
\end{figure}

To assess the validity of the bootstrap confidence intervals, this subsection compares the asymptotic confidence intervals for conditional EVaR calculated using \eqref{eq:ci} with the bootstrap confidence intervals obtained from \eqref{eq:ci_boot}.
The number of bootstrap replications is set to $M=1000$.
We consider two DGPs: GARCH model and GJR-GARCH model.
Figures~\ref{fig:cp_n500}--\ref{fig:cp_n1000} present the empirical coverage probabilities of the two methods at the 95\% confidence level across different $\tau$ levels for $n=500$ and $n=1000$, respectively.
It can be observed that in all settings, the empirical coverage probabilities of both methods fluctuate around the nominal 95\% level, indicating that the bootstrap procedure provides reliable confidence intervals for conditional EVaR.

\section{Empirical Illustration} 
\label{s:app}
In this section, we provide an empirical application to  S\&P 500 dataset by employing our proposed method in forecasting the EVaR of S\&P 500 returns.   
The data are 
based on the daily log returns of the S\&P 500 index over the ten-year period from January 4, 2016 to December 31, 2025, 
totally containing 2514 observations.  
The daily log return is defined as $Y_t = 100 \times (\log P_t - \log P_{t-1})$, where $P_t$ is the closing price of the S\&P 500 index on day $t$. 

\begin{figure}
\centering 
\includegraphics[width=0.9\linewidth]
{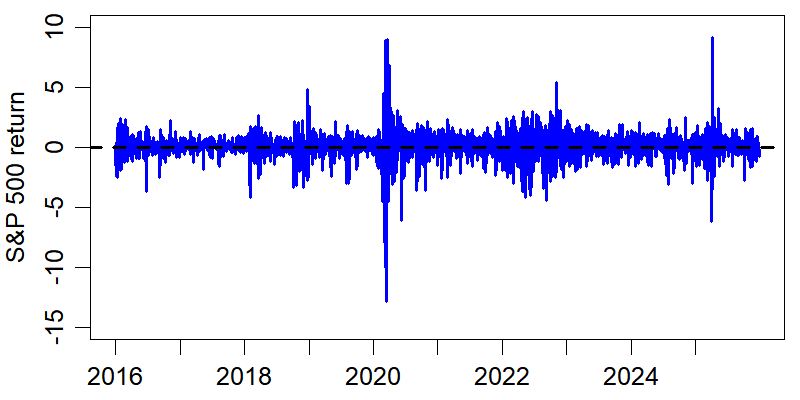}
\caption{Daily log returns of the S\&P500 index from January 4, 2016 to December 31, 2025.}
\label{fig:return}
\end{figure}

Figure~\ref{fig:return} presents the time series $\{Y_t\}$ and reveals pronounced volatility clustering. 
The descriptive statistics reported in Table~\ref{tab:description} indicate that the return distribution is sharply peaked with heavy tails. 
The augmented Dickey--Fuller (ADF) test strongly rejects the presence of a unit root, 
supporting stationarity. 
The Jarque--Bera (JB) test rejects the null hypothesis of normality, and the Lagrange Multiplier (LM) test provides strong evidence of ARCH effects. 
Motivated by these findings, 
it is reasonable to fit the data by  a GARCH$(1,1)$ model:
$ Y_{t}= \sigma_t \varepsilon_t, ~
\sigma_t^2 = \omega_0 + \omega_1 Y_{t-1}^2 + \omega_2\sigma_{t-1}^2$, 
and then conduct out-of-sample evaluation.

\begin{table}%[H]
\centering
\caption{Descriptive statistics of daily log returns for the S\&P500 index.}
\resizebox{\textwidth}{!}{
\begin{tabular}{cccccccccc}
\toprule
Mean & Med & Min & Max & SD & Skew & Kurtosis & ADF & JB & LM \\
\midrule
0.05 & 0.08 & -12.77 & 9.09 & 1.15 & -0.68 & 19.51 & -13.55$^{***}$ & 28810.12$^{***}$ & 813.45$^{***}$ \\
\bottomrule
\end{tabular}
}
\par\vspace{2mm} 
\noindent\footnotesize
\parbox{\textwidth}{\textit{Notes:} ADF, augmented Dickey--Fuller test statistic; JB, Jarque--Bera test statistic; LM, Lagrange Multiplier test statistic; $^{***}$ denotes being significant at the 1\% level.}
\label{tab:description}
\end{table}

We use the first 1514 observations as the training data for model estimation,
and reserve the remaining $n=1000$ observations from January 6, 2022 to December 31, 2025 for out-of-sample evaluation. 
For comparison, we also report results for the nonparametric estimator (``NPE'') by \citet{zhang2025nonparametric}, which exhibited relatively better performance among the two nonparametric methods considered in Subsection~\ref{ss:comparison}.

\begin{figure}
  \centering
  
  \begin{subfigure}{0.49\textwidth}
    \centering
    \includegraphics[width=\linewidth]{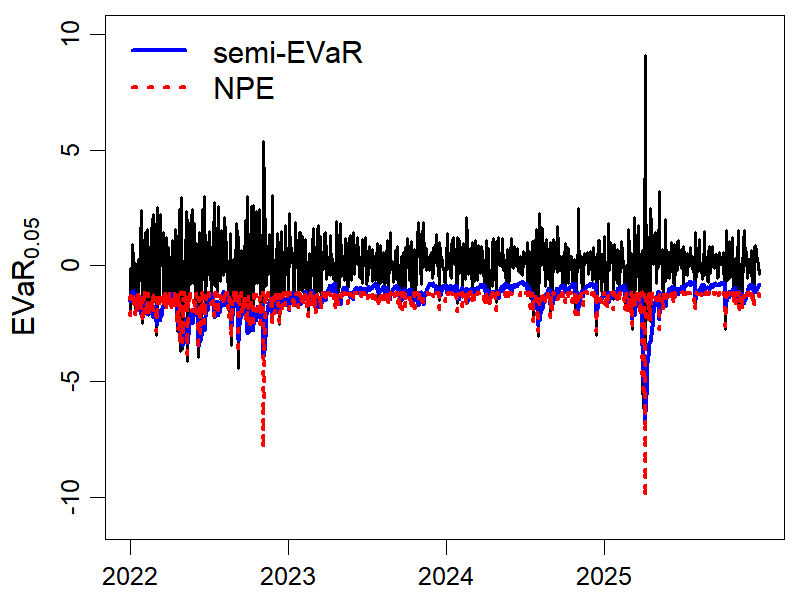}
    \caption{Estimated 5\% conditional EVaR}
  \end{subfigure}\hfill
  \begin{subfigure}{0.49\textwidth}
    \centering
    \includegraphics[width=\linewidth]{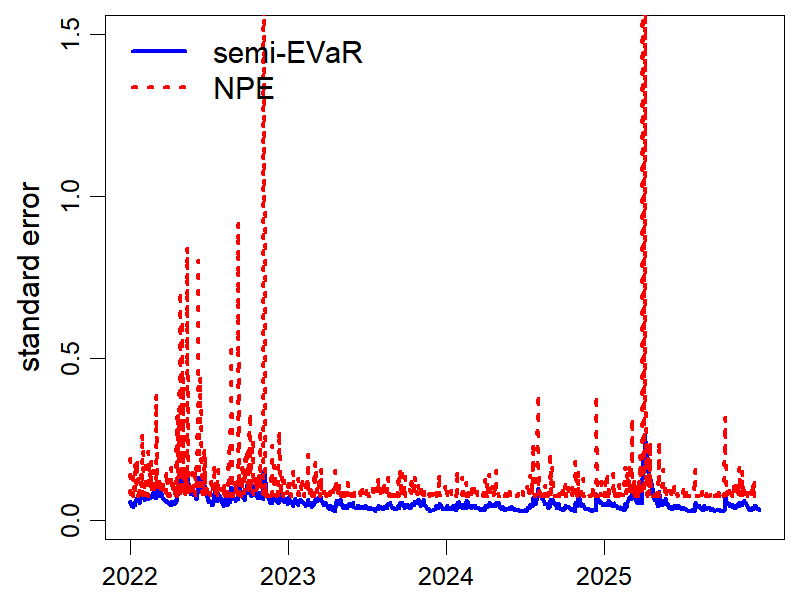}
    \caption{Estimated standard error}
  \end{subfigure}
  
  \caption{Estimated 5\% conditional EVaR and estimated standard errors for the last 1000 out-of-sample observations of the S\&P 500 index. 
  Black line: the actual daily log return; 
  blue line: the proposed semiparametric estimator (semi-EVaR); 
  red line: the nonparametric estimator by \citet{zhang2025nonparametric} (NPE).}
  \label{fig:evar_and_se}
\end{figure}

First, we estimate the 5\% conditional EVaR of $Y_t$ given 
$\bX_t=(Y_{t-1},Y_{t-2},\cdots)^\top$
based on the training data.   
Figure~\ref{fig:evar_and_se} displays the estimated 5\% conditional EVaR and estimated standard errors for the last 1000 out-of-sample observations using our proposed semiparametric estimator and the nonparametric estimator. 
Overall, both methods capture the main variation in downside risk. However, our estimator yields smaller and noticeably more stable estimated standard errors.

To evaluate EVaR forecast accuracy, as suggested by \cite{bellinirisk2017} and \cite{zhang2025nonparametric}, we consider the associated realized loss for out-of-sample prediction,
\[
\mathcal{L}(\tau)
= n^{-1}\sum_{t=1}^{n}\Psi_{\tau}\!\left(Y_t-\widehat{\operatorname{EVaR}}_{\tau}(X_t)\right),
\]
where $\Psi_\tau(\cdot)$ is the expectile loss defined in \eqref{eqn:evar}. 
Table~\ref{tab:loss} reports the realized loss $\mathcal{L}(\tau)$ for $\tau = 0.05, 0.10, 0.25, 0.50, 0.75, 0.90, 0.95$. It can be observed that our proposed estimator achieves smaller realized losses across all expectile levels.

\begin{table}%[H]
\centering
\caption{Realized loss $\mathcal{L}(\tau)$ for the S\&P 500 index. Smaller is better. }
\newcolumntype{Y}{>{\centering\arraybackslash}X}
\begin{tabularx}{\columnwidth}{l *{7}{Y}}
\toprule
$\tau$   & 0.05 & 0.10 & 0.25 & 0.50 & 0.75 & 0.90 & 0.95 \\
\midrule
semi-EVaR & \textbf{0.252} & \textbf{0.361} & \textbf{0.543} & \textbf{0.628} & \textbf{0.527} & \textbf{0.339} & \textbf{0.228} \\
NPE      & 0.255 & 0.370 & 0.550 & 0.633 & 0.540 & 0.361 & 0.254 \\
\bottomrule
\end{tabularx}
\label{tab:loss}
\end{table}

\begin{figure}
  \centering
  
  \begin{subfigure}{0.49\textwidth}
    \centering
    \includegraphics[width=\linewidth]{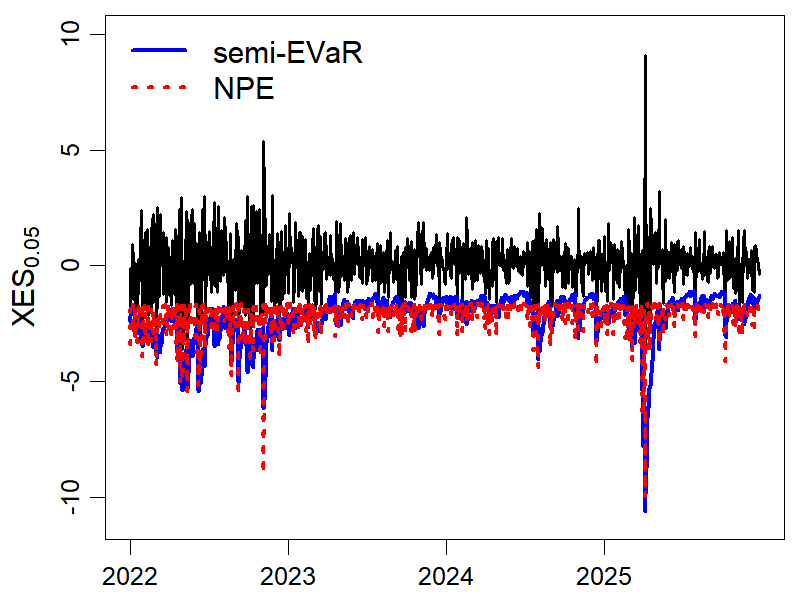}
    \caption{Estimated 5\% XES}
  \end{subfigure}\hfill
  \begin{subfigure}{0.49\textwidth}
    \centering
    \includegraphics[width=\linewidth]{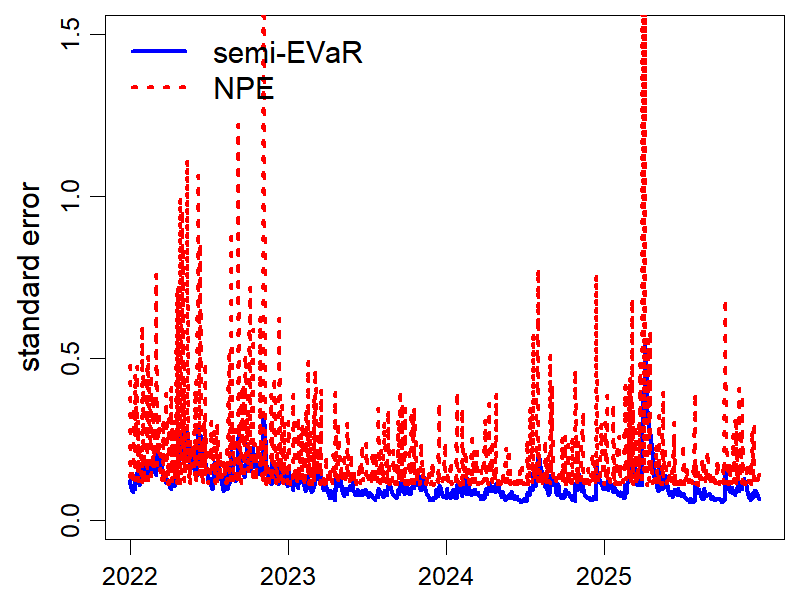}
    \caption{Estimated standard error}
  \end{subfigure}
  
  \caption{Estimated 5\% XES and estimated standard error for the last 1000 out-of-sample observations of the S\&P 500 index. 
  Black line: the actual daily log return; 
  blue line: the proposed semiparametric estimator (semi-EVaR); 
  red line: the nonparametric estimator by \citet{zhang2025nonparametric} (NPE).}
  \label{fig:xes_and_se}
\end{figure}

Next, we estimate the  conditional 
$\operatorname{XES}_{0.05}(X_t)$, 
with results presented in Figure~\ref{fig:xes_and_se}. 
Similar to the EVaR case, the proposed estimator delivers more stable uncertainty quantification, as evidenced by smaller estimated standard errors than NPE. 
Moreover, the estimated $\operatorname{XES}_{0.05}(X_t)$ is generally more negative 
compared with the corresponding $\operatorname{EVaR}_{0.05}(X_t)$ in Figure~\ref{fig:evar_and_se}, indicating that EVaR tends to be more conservative than XES at the same expectile level.

\begin{figure}
  \centering
  
  \begin{subfigure}{0.49\textwidth}
    \centering
    \includegraphics[width=\linewidth]{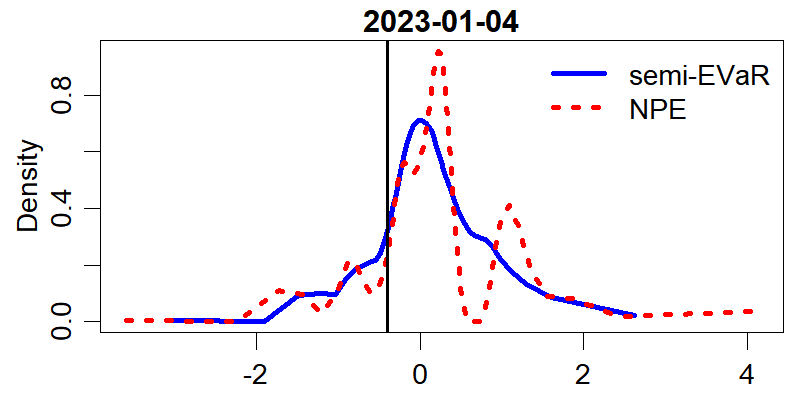}
  \end{subfigure}\hfill
  \begin{subfigure}{0.49\textwidth}
    \centering
    \includegraphics[width=\linewidth]{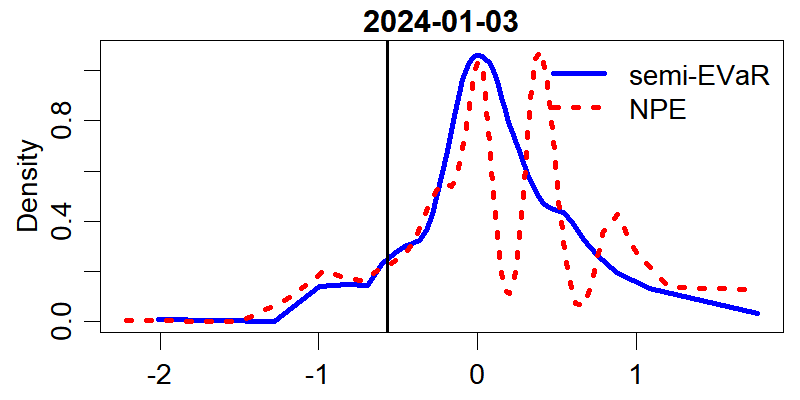}
  \end{subfigure}

  \vspace{0.5em}

  \begin{subfigure}{0.49\textwidth}
    \centering
    \includegraphics[width=\linewidth]{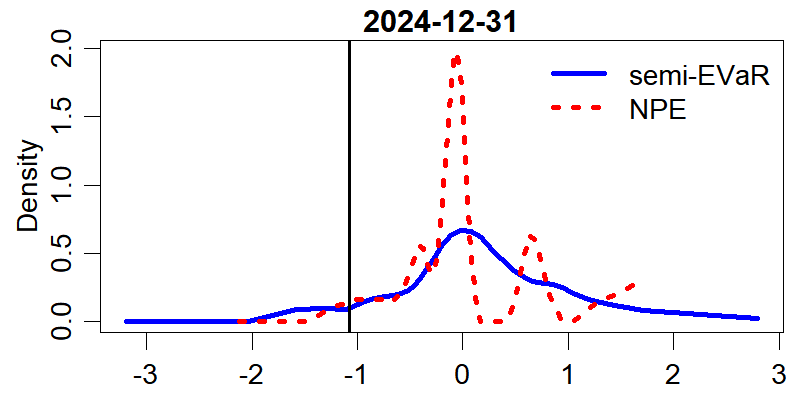}
  \end{subfigure}\hfill
  \begin{subfigure}{0.49\textwidth}
    \centering
    \includegraphics[width=\linewidth]{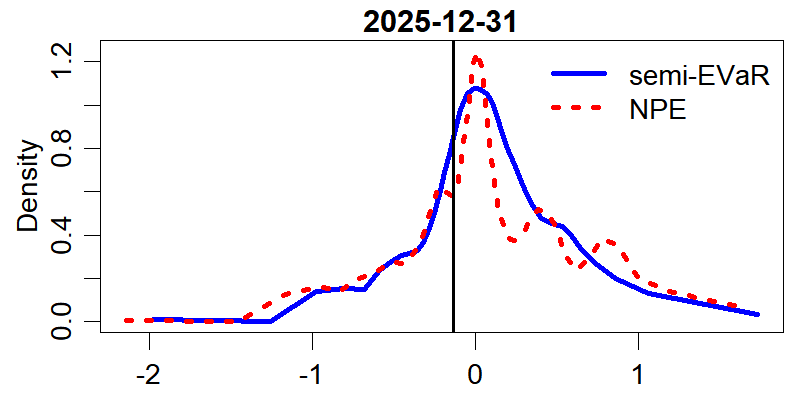}
  \end{subfigure}

  \caption{Predicting densities at four representative forecast dates. 
  Black line: the realized return on the forecast day; 
  blue line: the proposed semiparametric estimator (semi-EVaR); 
  red line: the nonparametric estimator by \citet{zhang2025nonparametric} (NPE).}
  \label{fig:density}
\end{figure}

Finally, we assess density forecasts. 
Figure~\ref{fig:density} depicts   
four representative one-step-ahead predictive densities corresponding to $n=250,500,750,1000$ (i.e., 2023-01-04, 2024-01-03, 2024-12-31, and 2025-12-31, respectively). 
Across the four subfigures, the proposed predictive density is smooth and predominantly unimodal, whereas the nonparametric density can be noticeably more irregular, exhibiting local oscillations around the center.

Density forecast accuracy is evaluated via the predictive log-score (LS) and the weighted expectile score (WES). 
The LS is defined by
$
\operatorname{LS}_{t} = -\log f_t(Y_t),
$
where $f_t(\cdot)$ denotes the predictive density evaluated at the realized return $Y_t$. 
Similar to the weighted quantile score in \cite{gneiting2011comparing}, the WES is defined as
\[
\operatorname{WES}_{t}=\int_{0}^{1} \Psi_{\tau}\!\left(Y_t-\widehat{\operatorname{EVaR}}_{\tau}(X_t)\right)\omega(\tau)d\tau,
\]
where $\omega(\tau)$ is a weight function on $(0,1)$. 
Lower LS and WES indicate better distributional forecasts.
Following \cite{busetti2017quantile}, we consider five choices:

\vspace{6pt}
\begin{tabular}{@{}ll@{}}
(1) uniform: $\omega(\tau)=1$;
& (2) center: $\omega(\tau)=\tau(1-\tau)$; \\[4pt]

(3) left tail: $\omega(\tau)=(1-\tau)^2$; 
& (4) right tail: $\omega(\tau)=\tau^2$; \\[4pt]

(5) double tails: $\omega(\tau)=(2\tau-1)^2$ .
&
\end{tabular}
\vspace{6pt}

To assess statistical significance in performance differences, we apply the Diebold--Mariano (DM) test, which is also used by \cite{hua2013forecasting}.
Specifically, we test the null hypothesis of equal predictive accuracy between the proposed method and the nonparametric estimator. 
We take NPE as the benchmark. 
Hence, a significantly negative DM statistic indicates that the proposed estimator achieves a lower average score and therefore outperforms NPE.

Table~\ref{tab:densityscore} summarizes the density forecast evaluation based on the LS and five variants of the WES.
For each criterion, we report the average scores of the two estimators and the DM test statistic, with the corresponding $p$-value shown in parentheses. 
Across all scoring rules, the proposed method achieves consistently smaller average scores than the NPE benchmark, indicating improved distributional forecasts. 
Moreover, the DM test results suggest that this improvement is statistically significant in most cases.

\begin{table}
\centering
\caption{Evaluation results of forecast distributions based on the predictive log-score (LS) and five versions of the weighted expectile score (WES).}
\newcolumntype{Y}{>{\centering\arraybackslash}X}
\begin{tabularx}{\columnwidth}{l *{6}{Y}}
\toprule
 & \multicolumn{1}{c}{$\overline{\mathrm{LS}}$} & \multicolumn{5}{c}{$\overline{\mathrm{WES}}$} \\
\cmidrule(lr){3-7}
 &  & uniform & center & left & right & both \\
\midrule
semi-EVaR & \textbf{1.701} & \textbf{0.489} & \textbf{0.548} & \textbf{0.474} & \textbf{0.444} & \textbf{0.365} \\
NPE      & 3.574 & 0.503 & 0.560 & 0.493 & 0.453 & 0.381 \\
\midrule
DM test  & -10.353 & -1.590 & -1.200 & -1.464 & -0.969 & -2.726 \\
 $p$-value        & (0.000) & (0.056) & (0.115) & (0.072) & (0.166) & (0.003) \\
\bottomrule
\end{tabularx}
\label{tab:densityscore}
\end{table}

\section{Conclusion} 
\label{s:con}
This paper develops semiparametric estimation and inference for conditional EVaR in nonlinear time series models with an unspecified error distribution. 
We propose a new two-step conditional EVaR estimator and establish its asymptotic properties. 
A residual bootstrap procedure is introduced for constructing confidence intervals.
We also investigate semiparametric estimation of the risk measure XES and the conditional distribution by the conditional EVaR process. 
Monte Carlo experiments and an empirical application to S\&P~500 returns confirms the superior performance of the proposed semiparametric estimator.
%Monte Carlo experiments indicate that the proposed estimator consistently outperforms two competitive nonparametric estimators and remains robust under model misspecification.
%An empirical application to S\&P~500 returns further confirms the strong predictive performance of our estimator for conditional EVaR, XES, and distribution forecasts.

There are several directions for future work. 
First, 
it would be interesting to extend the proposed framework to settings with high-dimensional predictors %by using regularization techniques %\citep{xu2021elastic} 
~\citep{gu2016HighDimensional}
or with nonparametric location-scale models by machine learning methods in the first-step estimation~\citep{zhao2026minimax}.
Another promising direction is to consider extreme conditional expectiles~\citep{girard2021extreme}. 
We leave these interesting topics as possible future work.

\iffalse
\section*{Funding}
%%The authors are grateful to the three anonymous referees for their helpful comments.  %The first two authors contributed equally to this work.
% To JoE, ET, Journal of Financial Econometrics. 
This work was financially supported by the National Natural Science Foundation of China (No. 72171192), RGC grant of Hong Kong Research Grant Council (No. 12302022, 12303926) and  Initiation Grant for Faculty Niche Research Areas of Hong Kong Baptist University (No. RC-FNRA-IG/24-25/SCI/01).

\section*{Competing Interests}
The authors declare that no competing interests exist.

\fi

%%%%%%%%%%%%%%%%%%%%%%%%%%%%%%%%%%%%%%%%%%%%%%%%%%%%%%%%%%%%%
%% the reference 
\begin{spacing}{1.3}
\bibliographystyle{apalike}
\bibliography{ER_ref}
\end{spacing}
%%%%%%%%%%%%%%%%%%%%%%%%%%%%%%%%%%%%%%%%%%%%%%%%%%%%%%%%%%%%%%%%%%%%%%	

\end{document}